\documentclass[11pt,a4paper]{article}
\pdfoutput=1
\usepackage{jheppub}
\usepackage{amsthm}
\usepackage{microtype} 
\usepackage{booktabs}
\usepackage{bm}
\usepackage{tikz}
\usetikzlibrary{positioning, arrows.meta}

\title{\boldmath Holographic entanglement entropy, Wilson loops, and neural networks}

\author[a]{Veselin G.~Filev\,\href{https://orcid.org/0000-0001-5786-6815}{\textsuperscript{\normalfont\tiny ORCID}}}

\affiliation[a]{Institute of Mathematics and Informatics, Bulgarian Academy of Sciences,\\
Acad.\ G.\ Bonchev Str., 1113 Sofia, Bulgaria}

\emailAdd{vfilev@math.bas.bg}

\abstract{We apply artificial neural networks to the holographic inverse problem, reconstructing bulk geometry from boundary entanglement entropy by using the Ryu--Takayanagi area functional as a differentiable loss. Validated on the AdS-Schwarzschild background, this approach recovers the blackening factor with maximum absolute error below $3\times10^{-3}$ across the entire bulk, reproducibly over independent training runs. For finite-density backgrounds like the Gubser--Rocha model, we demonstrate that equal-time strip entanglement entropy determines only the spatial metric. We resolve this exact one-function degeneracy by incorporating holographic Wilson loop data, which couples to the timelike metric. We present a semi-analytical inversion combining Bilson's and Hashimoto's formulas, alongside a general three-network variational method minimizing the combined area and Nambu--Goto actions. The neural network achieves maximum relative errors below $0.2\%$ for both metric functions without closed-form derivative relations, and accommodates additional holographic observables at the cost of one extra network and loss term.}

\makeatletter
\gdef\@fpheader{Published in \href{https://doi.org/10.1007/JHEP09(2026)073}{JHEP 09 (2026) 073}}
\makeatother
\arxivnumber{2604.05970}
\begin{document}

\maketitle
\flushbottom

\section{Introduction}
\label{sec:intro}

The Ryu--Takayanagi (RT) prescription~\cite{Ryu:2006bv,Ryu:2006ef} provides a concrete geometric realization of entanglement entropy in the AdS/CFT correspondence: the entanglement entropy $S_{EE}$ of a boundary subregion is given by the area of a bulk minimal surface anchored on the boundary of that region,
\begin{equation}\label{eq:RT}
  S_{EE}(A) = \frac{\text{Area}(\Gamma_A)}{4\, G_N}\,.
\end{equation}
For a strip subregion of width~$l$ in a $d$-dimensional CFT at finite temperature, the dual geometry is AdS-Schwarzschild and the problem reduces to finding the minimal area surface in the bulk.
Below a critical strip width $l_c$ the minimal surface is a connected surface dipping into the bulk, while above $l_c$ the disconnected surface (two half-planes hanging from the boundary to the horizon) has lower area.
The transition at $l_c$ is first order: beyond $l_c$ the finite part of the entanglement entropy saturates, reflecting the thermal screening of correlations at scales of order $1/T$.

Computing the minimal surface $\Gamma_A$ typically requires deriving non-linear Euler--Lagrange equations from the area functional and solving the resulting boundary-value problem.
In ref.~\cite{Filev:2025} it was shown that artificial neural networks (ANNs) can bypass this step entirely for the case of probe-brane embeddings: by parametrizing the embedding with an ANN and using the regularized DBI action as the loss function, the network learns the equilibrium profile purely through gradient descent --- no equation of motion is derived or solved explicitly.
The same work demonstrated that a conditional network can learn an entire one-parameter family of embeddings and that the bulk geometry can be reconstructed from boundary data via alternating optimization.

In this paper we extend this approach from probe-brane embeddings to RT surfaces, where the area functional plays the role of the DBI action.
Our goals are:
\begin{enumerate}
  \item To show that an ANN with the RT area functional as loss accurately reproduces the minimal surface profiles $z(x)$ and the entanglement entropy $S_{EE}(l)$ for strip subregions in AdS$_5$-Schwarzschild, including the connected/disconnected phase transition.
  \item To demonstrate that a conditional network $z(x,l)$ learns the full one-parameter family of surfaces and enables automatic differentiation of $S_{EE}$ with respect to~$l$.
  \item To solve the inverse problem: given entanglement entropy data $S_{EE}(l)$, recover the bulk metric. For metrics with $h(z) \neq 1$, we investigate the fundamental degeneracy in the data and provide two complementary resolutions using Wilson loop data.
\end{enumerate}
The third point is the paper's main contribution.\footnote{The code accompanying this paper is available at \url{https://github.com/vesofilev/holographic_ee_wl}.}
For the AdS-Schwarzschild case ($h = 1$), the inverse problem is well-posed, and we use it to validate the ANN against Bilson's semi-analytical inversion~\cite{Bilson:2008,Bilson:2010ff} before tackling the genuinely new problem.
For metrics with two unknown functions ($h \neq 1$), we prove that $S_{EE}(l)$ determines only the spatial metric component $g(r)$ and not the timelike component $\chi(r)$, implying a fundamental one-function degeneracy.
We resolve this degeneracy by supplementing entanglement entropy with holographic Wilson loop data, exploiting the fact that the string worldsheet extends in time and thereby probes~$\chi$.
We present two complementary reconstruction methods: a semi-analytical inversion combining Bilson's entanglement entropy formula~\cite{Bilson:2008,Bilson:2010ff} with Hashimoto's Wilson loop formula~\cite{Hashimoto:2020}, and a general three-network variational approach.
The generality of the ANN approach is its key advantage: it does not require closed-form derivative relations, making it straightforward to add new observables without new semi-analytical derivations.

The application of neural networks to holographic problems has a growing literature.
Hashimoto et al.~\cite{Hashimoto:2018} pioneered the use of deep learning for bulk reconstruction in AdS/CFT.
Park et al.~\cite{Park:2022} reconstructed dual geometries from entanglement entropy using deep learning and the RT formula.
Ahn et al.~\cite{Ahn:2024} first addressed the neural-network inverse problem for entanglement entropy with two unknown metric functions, addressing the Gubser--Rocha and superconductor backgrounds using neural ODEs.
Kim~\cite{Kim:2025} uses a Transformer architecture trained on synthetic (geometry, $S_{EE}$) pairs to learn the inverse RT map.
On the analytic side, Jokela et al.~\cite{Jokela:2025} derive an inversion formula relating entanglement entropy variations to metric deviations, while Fan and Yang~\cite{Fan:2024,Fan:2026} reconstruct the bulk metric from boundary two-point correlation functions using inverse scattering methods.
Deb and Sanghavi~\cite{Deb:2025} use physics-informed neural networks (PINNs) to solve the Euler--Lagrange equations of the area functional; their method still requires deriving the equations of motion.

Our approach differs from the above methods in a specific way: we use the area functional \emph{directly} as a differentiable loss function, avoiding the explicit derivation and solution of the bulk Euler--Lagrange equations.
The neural network is a variational tool, not a differential-equation solver.
The prior structure built into our network ansatz is purely kinematic --- the near-boundary behavior shared by every geometry in the asymptotically AdS class, together with thermodynamic data of the state --- so that all information about the interior geometry is supplied by the boundary data.\footnote{For example, the endpoint exponent $1/d$ in the boundary-condition encoding of section~\ref{sec:bc} follows from the conserved Hamiltonian of $x$-translations evaluated near the boundary, where every admissible metric approaches pure AdS; it is therefore identical for all candidate geometries and carries no information about the functions being reconstructed; see appendix~\ref{app:asymptotics} for a discussion from both the bulk and the CFT perspectives.}
The distinction is sharpest against PINN methods~\cite{Deb:2025} that solve the Euler--Lagrange equations; relative to the neural-ODE approach of ref.~\cite{Ahn:2024}, which integrates the turning-point parametrized integrals, the difference lies in the fully variational philosophy and the avoidance of the turning-point reduction.
Both the surface and the metric are learned simultaneously through the area functional via alternating optimization, extending the DBI framework of ref.~\cite{Filev:2025}.

The paper is organized as follows.
In section~\ref{sec:setup} we introduce the background geometry and area functional.
Section~\ref{sec:ann} describes the ANN architecture, boundary condition encoding, and training procedure, and presents results for single strip widths and the phase transition.
In section~\ref{sec:conditional} we extend to a conditional network that learns the full $S_{EE}(l)$ curve.
Section~\ref{sec:inverse} tackles the inverse problem: Section~\ref{sec:inverse_h1} validates the method on AdS-Schwarzschild, while section~\ref{sec:inverse_gr} analyses the Gubser--Rocha model, proves the metric degeneracy, and introduces the Wilson loop resolution.
Section~\ref{sec:noise} quantifies the robustness of the reconstruction with respect to the random seed and to noise in the boundary data.
Section~\ref{sec:comparison} compares the ANN and ODE methods.
We conclude in section~\ref{sec:conclusion}.

\section{RT surfaces in AdS-Schwarzschild}
\label{sec:setup}

We work throughout in the large-$N$, strong-coupling regime where the RT prescription applies and the minimal surface does not backreact on the geometry.
We consider the AdS$_5$-Schwarzschild background in Fefferman--Graham coordinates:
\begin{equation}\label{eq:metric}
  ds^2 = \frac{1}{z^2}\left[-f(z)\, dt^2 + dx_1^2 + dx_2^2 + dx_3^2 + \frac{dz^2}{f(z)} \right],
\end{equation}
where
\begin{equation}
  f(z) = 1 - \left(\frac{z}{z_h}\right)^4
\end{equation}
is the blackening factor.
The boundary is at $z = 0$ and the horizon at $z = z_h$; the Hawking temperature is $T = d/(4\pi z_h)$ with $d = 4$.
We set the AdS radius $L = 1$ and work in units where $z_h = 1$.

For a strip subregion $x_1 \in [-l/2,\, l/2]$ with transverse volume $V_2$, we parametrize a static RT surface by $z = z(x)$ ($x \equiv x_1$).
The induced metric yields the area functional
\begin{equation}\label{eq:area}
  A = V_2 \int_{-l/2}^{l/2} dx\; \frac{1}{z^3}\,\sqrt{1 + \frac{z'^2}{f(z)}}\,.
\end{equation}
The Lagrangian $\mathcal{L}(z,z') = z^{-3}\sqrt{1 + z'^2/f}$ does not depend explicitly on~$x$, so the Hamiltonian
\begin{equation}\label{eq:firstintegral}
  H = \frac{1}{z^3\,\sqrt{1 + z'^2/f(z)}} = \frac{1}{z_*^3}
\end{equation}
is a first integral of the Euler--Lagrange equations, where $z_*$ is the turning point at $x = 0$.
This first integral is the basis of the traditional ODE shooting method and will be used to generate benchmark data; the ANN approach does not invoke it.

The regularized area is defined as $A_{\text{reg}} = A_{\text{conn}} - A_{\text{disc}}$, where $A_{\text{disc}}$ is the area of the disconnected surface --- two straight embeddings at $x = \pm l/2$ extending from $z = \epsilon$ to the horizon:
\begin{equation}\label{eq:Adisc}
  A_{\text{disc}} = 2V_2 \int_{\epsilon}^{z_h} \frac{dz}{z^3\,\sqrt{f(z)}}\,.
\end{equation}
The entanglement entropy is $S_{EE} = \min(A_{\text{conn}},\, A_{\text{disc}})/(4G_N)$.
At the critical width $l_c \approx 0.888\, z_h$, the two areas are equal and the system undergoes a first-order transition.

\section{Extremising the area functional with neural network}
\label{sec:ann}

Our first goal is to construct an ANN that learns the RT surface profile $z(x)$ for a single strip width~$l$ by minimizing the regularized area functional.
This extends the approach of ref.~\cite{Filev:2025} from the DBI action to the RT area functional.

\subsection{Network architecture}
\label{sec:architecture}

\begin{figure}[ht]
\centering
\begin{tikzpicture}[
  neuron/.style={circle, draw, minimum size=18pt, inner sep=0pt, font=\small},
  >=Stealth
]
\node[neuron, fill=green!20] (input) at (0,0) {$x^2$};
\foreach \l/\x in {1/2.5, 2/5} {
  \node[neuron, fill=blue!10] (h\l a) at (\x, -1.0) {};
  \node[neuron, fill=blue!10] (h\l b) at (\x, -0.33) {};
  \node[neuron, fill=blue!10] (h\l c) at (\x, 0.33) {};
  \node[neuron, fill=blue!10] (h\l d) at (\x, 1.0) {};
  \node[above] at (\x, 1.5) {\small Layer \l};
  \node[below] at (\x, -1.5) {\small tanh};
}
\node[neuron, fill=orange!20] (output) at (7.5, 0) {$g$};
\foreach \n in {a,b,c,d} { \draw[->] (input) -- (h1\n); }
\foreach \a in {a,b,c,d} { \foreach \b in {a,b,c,d} {
  \draw[->, gray!40] (h1\a) -- (h2\b); }}
\foreach \n in {a,b,c,d} { \draw[->] (h2\n) -- (output); }
\node[below=0.3cm] at (input) {\small Input};
\node[below=0.3cm] at (output) {\small Output};
\node[neuron, fill=red!15, minimum size=24pt] (zout) at (10, 0) {$z(x)$};
\draw[->, thick] (output) -- node[above, font=\footnotesize] {eq.~\eqref{eq:bc}} (zout);
\end{tikzpicture}
\caption{Architecture of the ANN for the RT surface profile.
The input is $x^2$ (enforcing $z(-x) = z(x)$), the output is $g_{\text{NN}}(x^2)$, and the physical profile is obtained via the boundary condition encoding~\eqref{eq:bc}.
The actual network has 20~neurons per hidden layer.}
\label{fig:architecture}
\end{figure}
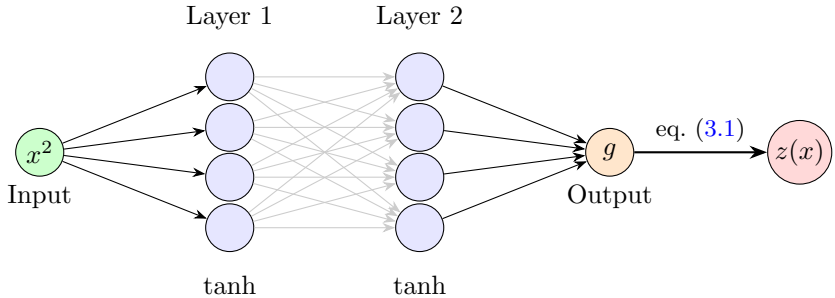

The ANN is a feedforward network with a single input $x^2$ and scalar output $g_{\text{NN}}(x^2)$.
It consists of two hidden layers of 20~neurons each (461~parameters total), with hyperbolic tangent activations.
The final output layer is linear.
The architecture is shown in figure~\ref{fig:architecture}.

\subsection{Boundary condition encoding}
\label{sec:bc}

To impose the boundary conditions we perform a final algebraic transformation.
Inspired by the exact RT surface in pure AdS ($f = 1$), where $z(x) \propto (l^2/4 - x^2)^{1/d}$ near the boundary, we use the ansatz
\begin{equation}\label{eq:bc}
  z_{\text{NN}}(x) = \epsilon + \left(\frac{l^2}{4} - x^2\right)^{\!1/d}\, \text{softplus}\!\left(g_{\text{NN}}(x^2)\right).
\end{equation}
This encoding satisfies:
\begin{itemize}
  \item $z'(0) = 0$ --- automatic from the $x^2$ input, making $z(x)$ even;
  \item $z(l/2) = \epsilon$ --- exact, from the vanishing prefactor;
  \item $z(0) = z_*$ --- free, learned by minimizing the area.
\end{itemize}
The exponent $1/d$ gives $z \sim (l/2 - x)^{1/d}$ near the boundary, so $z'(x) \to \infty$ as $x \to l/2$.
The softplus factor deforms the profile away from the pure-AdS shape to accommodate the blackening factor.
Although the encoding is motivated by the pure-AdS solution, the exponent $1/d$ is universal rather than solution-specific: near its anchor the surface probes only the $z \to 0$ region, where every metric of the class~\eqref{eq:metric} approaches pure AdS, so the same endpoint behavior holds for every candidate geometry (appendix~\ref{app:asymptotics}).
The encoding thus fixes structure common to the entire search space; all dynamical information about the interior remains to be learned.

\subsection{UV regularization of the loss}
\label{sec:reg}

Both $A_{\text{conn}}$ and $A_{\text{disc}}$ diverge as $1/\epsilon^2$, making their direct numerical subtraction unreliable.
The standard approach uses the first integral~\eqref{eq:firstintegral} to obtain a manifestly finite formula, but this invokes the equation of motion.
We regularize instead by re-parametrizing the near-boundary region in terms of the radial coordinate~$z$.

Using the symmetry $z(x) = z(-x)$, we split the connected half-area at $x_s = l/5$ and change variables from $x$ to $z$ in the boundary region $[x_s,\, l/2]$, where $x'(z) = 1/z'(x)$.
Since the disconnected surface is also a $z$-integral, the two integrands share the same $1/(z^3\sqrt{f})$ leading divergence and can be subtracted pointwise.
The result is
\begin{equation}\label{eq:Areg}
  A_{\text{reg,half}} = \int_0^{x_s}\! \frac{dx}{z^3}\sqrt{1 + \frac{z'^2}{f}}
  \;+\; \int_{\epsilon}^{z_{\text{mid}}}\! \frac{dz}{z^3}\!\left[\sqrt{x'^2 + \frac{1}{f}} - \frac{1}{\sqrt{f}}\right]
  - \int_{z_{\text{mid}}}^{z_h}\! \frac{dz}{z^3\sqrt{f}}\,,
\end{equation}
where $z_{\text{mid}} = z(x_s)$.
All three integrals are individually finite.
The split point $x_s$ is arbitrary and drops out of the final result.
Because the divergences cancel at the integrand level, the cutoff can be taken as small as $\epsilon = 10^{-4}$ without catastrophic cancellation.

The training loss is $\text{Loss} = A_{\text{reg,half}}(z_{\text{NN}})$, evaluated via the trapezoidal rule in $x$ (interior) and $z$ (boundary), with the remainder computed by quadrature.

\subsection{Results for single strip widths}
\label{sec:single_l_results}

\begin{figure}[ht]
\centering
\includegraphics[width=0.85\textwidth]{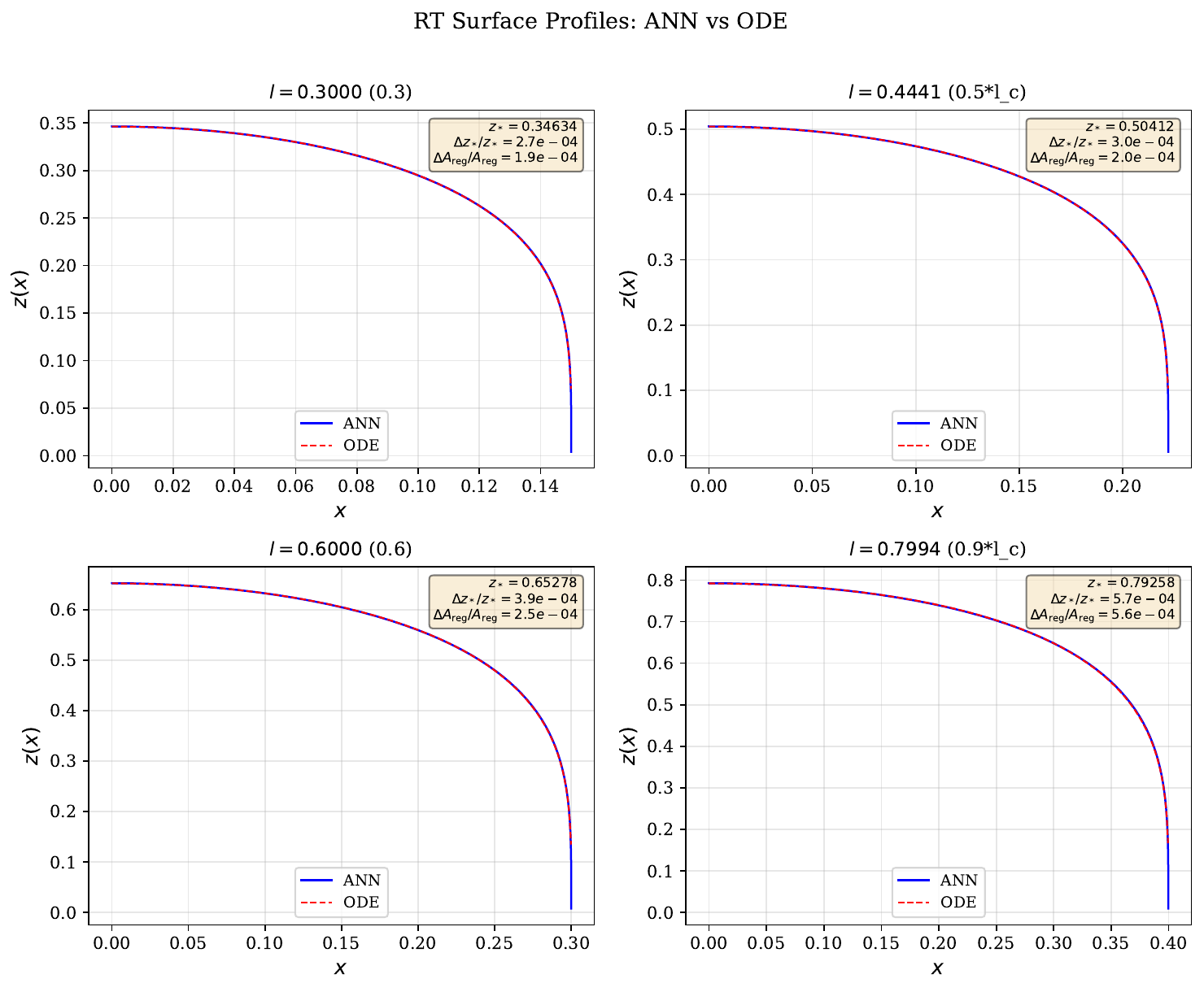}
\caption{RT surface profiles $z(x)$ from the ANN (solid) and the ODE shooting method (dashed) for four strip widths.
The ANN discovers the correct profile purely by area minimization.}
\label{fig:profiles}
\end{figure}

\begin{figure}[ht]
\centering
\includegraphics[width=0.85\textwidth]{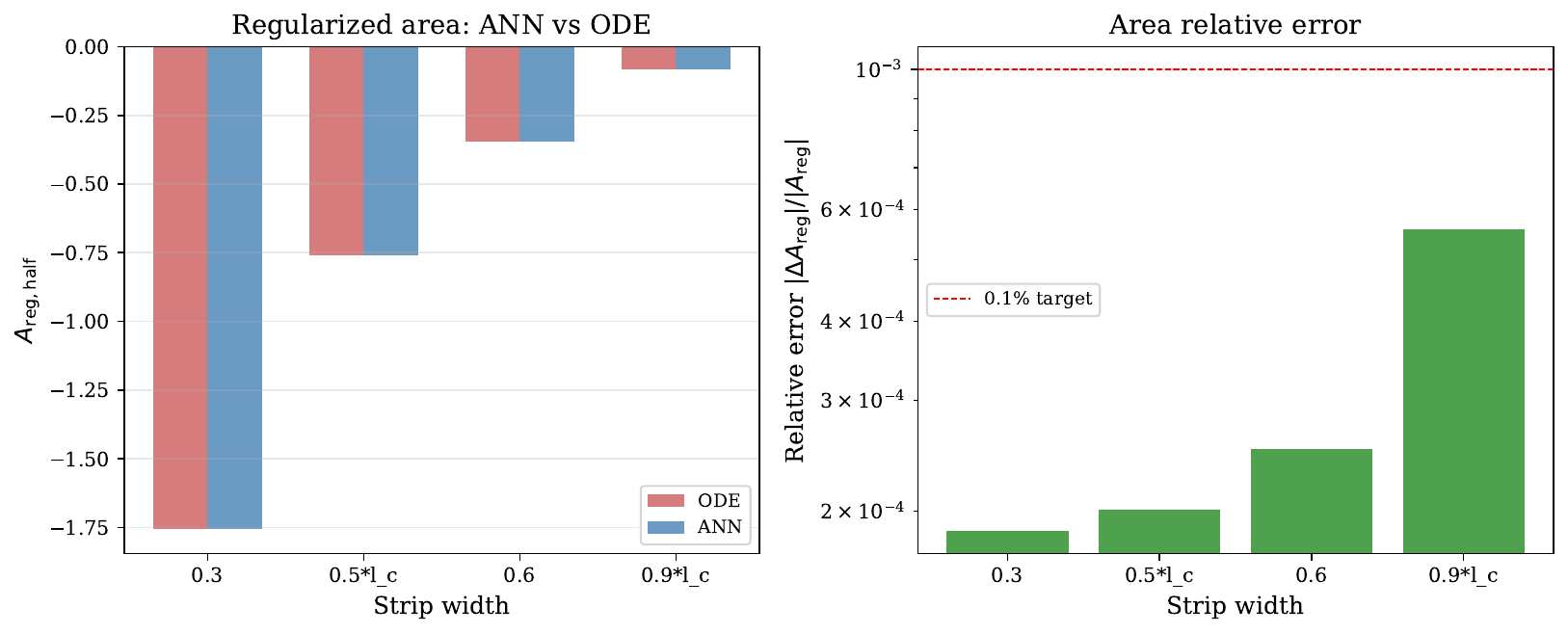}
\caption{Left: regularized area $A_{\text{reg,half}}$ from ODE and ANN for the four strip widths.
Right: relative area error, with the dashed line indicating the 0.1\% target.
All strip widths meet the target.}
\label{fig:area}
\end{figure}

\begin{figure}[ht]
\centering
\includegraphics[width=0.6\textwidth]{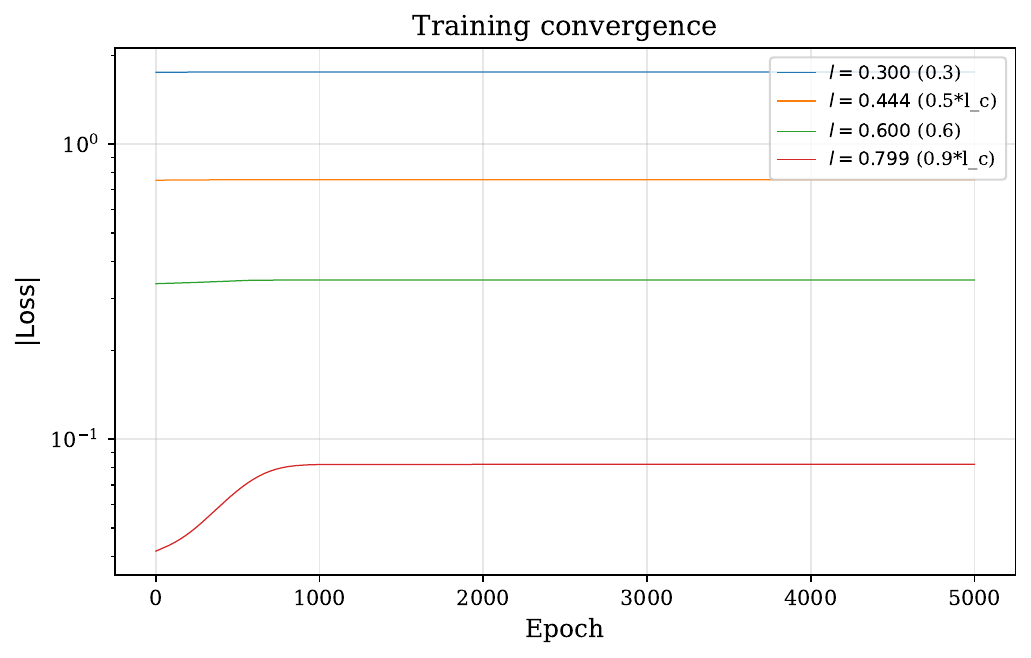}
\caption{Training convergence for the four strip widths.
The loss (regularized half-area) converges within ${\sim}\,1000$ epochs for all cases.}
\label{fig:convergence}
\end{figure}

We train the network for four strip widths spanning the range from narrow strips ($l = 0.3\,z_h$) to strips near the phase transition ($l = 0.9\,l_c$), using $\epsilon = 10^{-4}$, 5\,000 epochs, and a network with two hidden layers of 20~neurons (461 parameters).
Training takes approximately 35 seconds per strip width on a single CPU core.
The results are summarized in table~\ref{tab:single_l} and figures~\ref{fig:profiles}--\ref{fig:convergence}.

\begin{table}[ht]
\centering
\begin{tabular}{lcccccc}
\toprule
$l/z_h$ & $z_*^{\text{ODE}}$ & $z_*^{\text{ANN}}$ & $A_{\text{reg}}^{\text{ODE}}$ & $A_{\text{reg}}^{\text{ANN}}$ & $\delta z_*/z_*$ & $\delta A/A$ \\
\midrule
0.300 & 0.3462 & 0.3463 & $-1.7561$ & $-1.7564$ & $2.7\times 10^{-4}$ & $1.9\times 10^{-4}$ \\
$0.5\,l_c$ & 0.5040 & 0.5041 & $-0.7577$ & $-0.7578$ & $3.0\times 10^{-4}$ & $2.0\times 10^{-4}$ \\
0.600 & 0.6525 & 0.6528 & $-0.3462$ & $-0.3463$ & $3.9\times 10^{-4}$ & $2.5\times 10^{-4}$ \\
$0.9\,l_c$ & 0.7921 & 0.7926 & $-0.0821$ & $-0.0822$ & $5.7\times 10^{-4}$ & $5.6\times 10^{-4}$ \\
\bottomrule
\end{tabular}
\caption{Comparison of ANN and ODE results for four strip widths.
The relative errors in the turning point ($\delta z_*/z_*$) and regularized area ($\delta A/A$) are below $0.06\%$ in all cases.}
\label{tab:single_l}
\end{table}

The ANN reproduces the ODE benchmark to better than $0.06\%$ in all cases, without solving the equation of motion at any stage.
The accuracy degrades slightly for wider strips ($l \to l_c$), where the surface dips closer to the horizon and the profile has larger gradients.
The residual ${\sim}\,0.03\%$ error is consistent with the systematic effect of the finite UV cutoff $\epsilon = 10^{-4}$, which is absent in the ODE benchmark (computed via the $\epsilon$-independent first-integral formula).

\subsection{Phase transition}
\label{sec:phase_transition_single}

To map the connected/disconnected phase transition we train separate networks at 30~strip widths spanning $l \in [0.4\,l_c,\, 1.05\,l_c]$, with denser sampling near~$l_c$.
For each successive~$l$, the network is warm-started from the previous (nearby) solution.
This branch-tracking strategy --- already employed in section~2.2 of ref.~\cite{Filev:2025} for the meson-melting transition --- keeps the optimizer on the connected branch even for $l > l_c$, where the connected surface is metastable (higher area than the disconnected one, but still a local minimum of the area functional).

\begin{figure}[ht]
\centering
\includegraphics[width=0.9\textwidth]{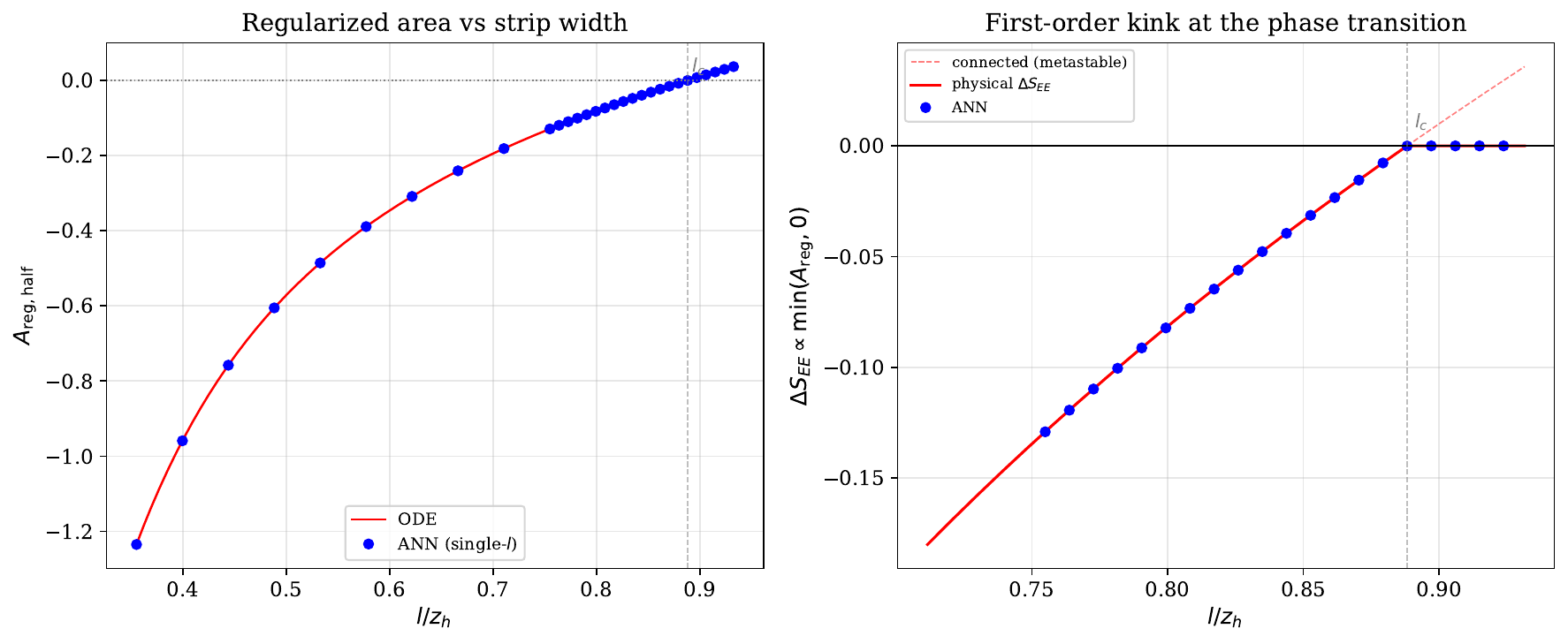}
\caption{Left: regularized area $A_{\text{reg,half}}$ as a function of strip width.
The ANN (blue circles) matches the ODE curve (red) across the full range, including the sign change at $l_c$.
Right: zoom near $l_c$ showing the physical $\Delta S_{EE} \propto \min(A_{\text{reg}}, 0)$ (solid red), which follows the connected branch for $l < l_c$ and vanishes for $l > l_c$.
The dashed line is the metastable connected branch past~$l_c$.
The kink at $l_c$ confirms the first-order nature of the transition.}
\label{fig:phase_transition}
\end{figure}

The results are shown in figure~\ref{fig:phase_transition}.
The ANN-computed $A_{\text{reg}}(l)$ matches the ODE curve across the full range, including the sign change at~$l_c$.
From a linear interpolation of the ANN data, the phase transition occurs at $l_c^{\text{ANN}} = 0.8883\,z_h$, compared with the ODE value $l_c = 0.8882\,z_h$ --- a relative error of $9 \times 10^{-5}$.
The network tracks the connected branch into the metastable region ($l > l_c$), where $A_{\text{reg}} > 0$ but the connected surface remains a local area minimum.

The physical entanglement entropy is $S_{EE}(l) \propto \min(A_{\text{conn}},\, A_{\text{disc}})$.
Since $A_{\text{disc}}$ is independent of~$l$, the finite part $\Delta S_{EE} \propto \min(A_{\text{reg}}, 0)$ has a kink at $l_c$: it follows the connected branch for $l < l_c$ and is identically zero for $l > l_c$ (right panel of figure~\ref{fig:phase_transition}).
The first derivative $d(\Delta S_{EE})/dl$ jumps from ${\approx}\,0.86$ to zero at $l_c$, confirming the first-order nature of the transition.

This demonstrates that the single-$l$ approach with warm-starting is well suited for studying phase transitions.
In section~\ref{sec:conditional} we compare this with a conditional network $z(x,l)$ that learns the full family simultaneously.

\section{Learning a one-parameter family of surfaces}
\label{sec:conditional}

Training a separate network for each strip width~$l$ is suboptimal: one expects the network weights to change smoothly with~$l$.
Following section~3 of ref.~\cite{Filev:2025}, we extend the architecture to a conditional network with two inputs $(x^2,\, l)$ and output $g_{\text{NN}}(x^2, l)$, so that a single network learns the entire family of surfaces $z(x,l)$.

The boundary condition encoding becomes
\begin{equation}\label{eq:bc_cond}
  z_{\text{NN}}(x, l) = \epsilon + \left(\frac{l^2}{4} - x^2\right)^{\!1/d}\, \text{softplus}\!\left(g_{\text{NN}}(x^2,\, l)\right).
\end{equation}
The loss is the average regularized area over a mini-batch of sampled strip widths.
Because $l$ is an input variable, the trained network provides direct access to the derivative $\partial_l S_{EE}$ via a single call to automatic differentiation.

\subsection{Results}

We train the conditional network for 60\,000 epochs on $l \in [0.1,\, 1.05\,l_c]$, sampling one strip width per optimization step (with 20\% probability of injecting the extremal values $l_{\min}$ or $l_{\max}$).
Note that, unlike the meson-melting set-up of ref.~\cite{Filev:2025} where the embedding can spontaneously change topology (Minkowski vs black hole), here the boundary condition encoding~\eqref{eq:bc_cond} forces the surface to be connected for all~$l$.
The disconnected surface is a completely different topology that the ansatz cannot produce.
We can therefore safely train across~$l_c$ and into the metastable region ($l > l_c$), where the connected surface still exists as a local area minimum.

Training takes ${\sim}\,28$~minutes on a single CPU core with 120\,000 epochs.
The network has two hidden layers of 30~neurons (1\,021~parameters).

\begin{figure}[ht]
\centering
\includegraphics[width=\textwidth]{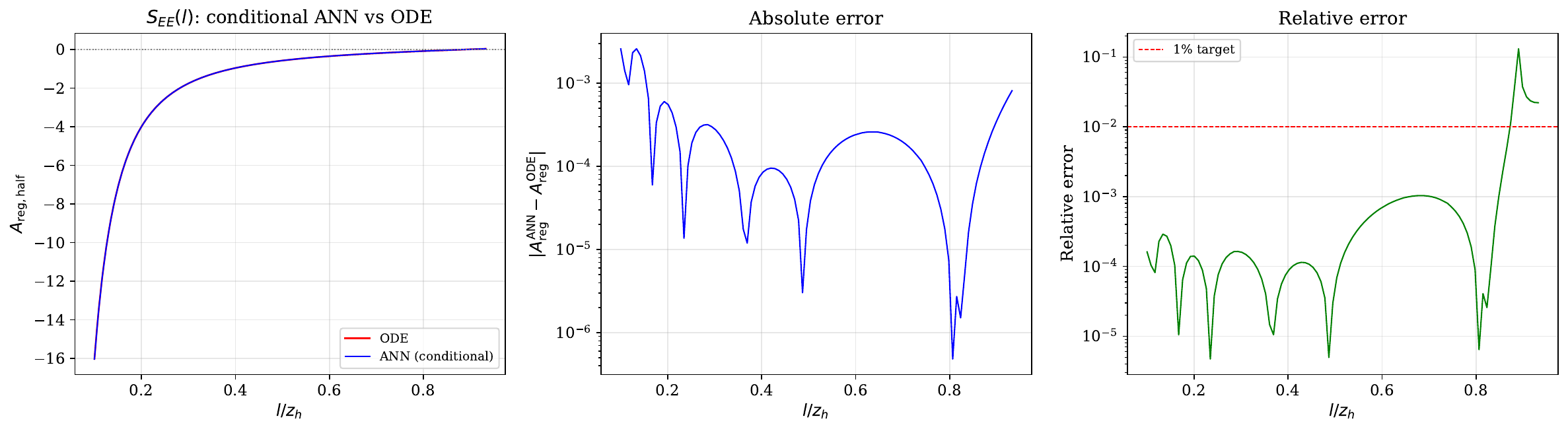}
\caption{Left: $A_{\text{reg,half}}(l)$ from the conditional ANN (blue) and ODE benchmark (red), including the metastable region $l > l_c$.
Center: absolute error $|A_{\text{reg}}^{\text{ANN}} - A_{\text{reg}}^{\text{ODE}}|$.
Right: relative error; the spike near $l_c$ is an artifact of $A_{\text{reg}} \to 0$.}
\label{fig:see_conditional}
\end{figure}

Figure~\ref{fig:see_conditional} shows the $S_{EE}(l)$ curve from the conditional network compared with the ODE benchmark.
Away from the phase transition (where $|A_{\text{reg}}| > 0.05$), the mean relative error is $0.03\%$ with a maximum of $0.10\%$.
Near $l_c$ the relative error formally diverges because $A_{\text{reg}} \to 0$, but the absolute error remains uniformly small (${\sim}\,3\times 10^{-4}$) across the full range (center panel).
The phase transition is identified at $l_c^{\text{ANN}} = 0.8879$, compared with the ODE value $l_c = 0.8882$ --- a relative error of $3\times 10^{-4}$.

\begin{figure}[ht]
\centering
\includegraphics[width=0.55\textwidth]{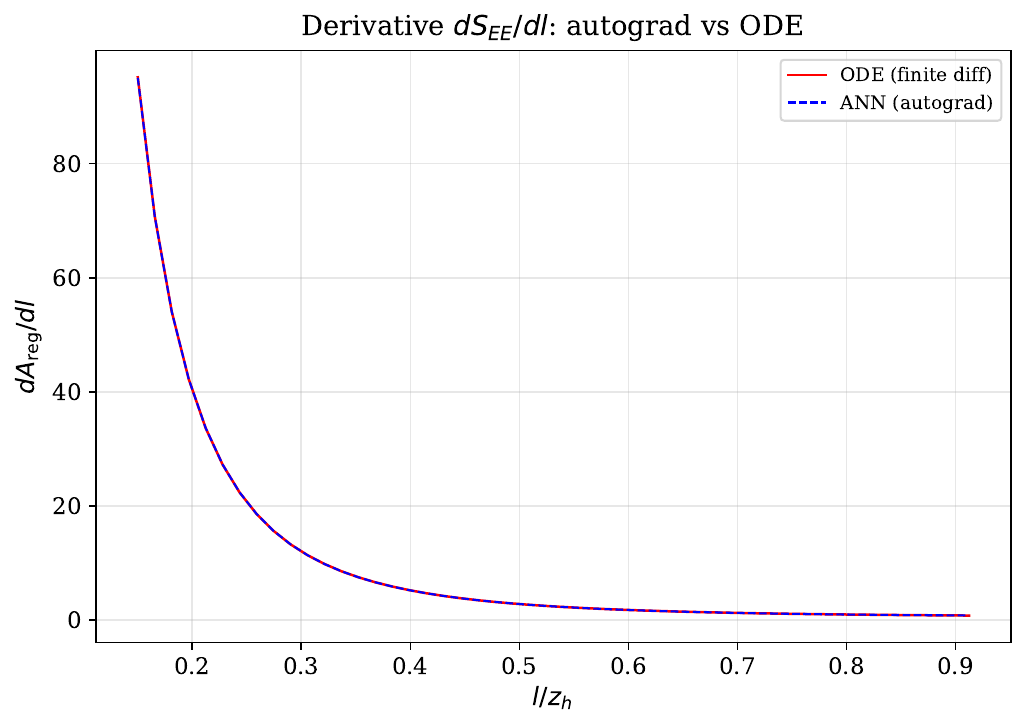}
\caption{Derivative $dA_{\text{reg}}/dl$ from the conditional ANN (dashed blue) and ODE finite differences (solid red).
The agreement demonstrates that the conditional network learns a smooth family of surfaces.}
\label{fig:condensate}
\end{figure}

Because $l$ is an input variable, the derivative $dS_{EE}/dl$ can be computed directly by finite-differencing the trained network at negligible cost.
Figure~\ref{fig:condensate} compares this with the ODE finite-difference derivative.

Compared with the single-$l$ approach of section~\ref{sec:phase_transition_single}, the conditional network trades a small loss in accuracy (0.16\% mean over all~$l$, versus ${\sim}\,0.03\%$ mean away from~$l_c$ for single-$l$ runs) for a large gain in efficiency: one network replaces 30+ separate trainings and provides the derivative for free.
For the inverse problem (section~\ref{sec:inverse}), the conditional network provides the warm-start and the differentiable $S_{EE}(l)$ functional needed for alternating optimization.

\section{Inverse problem: learning the geometry}
\label{sec:inverse}

In this section we address the central question: given entanglement entropy data $S_{EE}(l)$, can we reconstruct the bulk geometry?

For metrics with a single unknown function --- such as AdS-Schwarzschild, where $h(z) = 1$ --- the inverse problem admits an exact semi-analytical solution via Bilson's inversion formula~\cite{Bilson:2008,Bilson:2010ff}.
In the coordinate $r$ where the spatial metric takes the form $ds^2_{\text{spatial}} = r^{-2}[dr^2/g(r) + dx_i^2]$, Bilson showed that the entanglement entropy $S_{EE}(l)$ uniquely determines $g(r)$ through an Abel-type integral equation; for $h = 1$ this amounts to a complete reconstruction of~$f(z)$.
The ANN results presented in section~\ref{sec:inverse_h1} below serve as a validation of our variational method against this semi-analytical benchmark.
For metrics with $h(z) \neq 1$, however, Bilson's formula determines only one combination of the two unknown metric functions, and the inverse problem becomes fundamentally degenerate --- a point we analyse in detail in section~\ref{sec:inverse_gr}.

Concretely, we begin by recovering the blackening factor $f(z)$ in the AdS-Schwarzschild metric~\eqref{eq:metric} from the $S_{EE}(l)$ curve alone.
Following section~4 of ref.~\cite{Filev:2025}, we parametrize both the surface $z(x,l)$ and an unknown potential encoding the metric by separate ANNs.
The area functional~\eqref{eq:area} depends on $f(z)$ through the integrand, so the $S_{EE}(l)$ data constrains both the surface and the metric simultaneously.

We employ an alternating optimization scheme with two networks:
\begin{itemize}
  \item \textbf{L-model} (surface): a conditional network $z(x,l)$ with the same boundary condition encoding~\eqref{eq:bc_cond} as in section~\ref{sec:conditional}, with two hidden layers of 16~neurons (337~parameters).
  \item \textbf{V-model} (metric): a network encoding of $f_{\text{NN}}(z)$ with two hidden layers of 20~neurons (481~parameters), satisfying $f(0) = 1$ and $f(z_h) = 0$ by construction.
\end{itemize}
The V-model parametrizes the blackening factor as
\begin{equation}\label{eq:Vnorm}
  f_{\text{NN}}(z) = 1 + \zeta^2\bigl[-1 + D(\zeta) - D(1)\bigr]\,, \qquad \zeta \equiv z/z_h\,,
\end{equation}
where $D$ is the network output.
The encoding satisfies $f(0) = 1$ (asymptotically AdS) and $f(z_h) = 0$ (horizon) identically.
The absence of a linear term is boundary data in the sense of appendix~\ref{app:asymptotics}: the dual theory is an undeformed CFT, so no operator that could source a linear mode is turned on, and $f'(0) = 0$.
This is in contrast with the Gubser--Rocha problem of section~\ref{sec:inverse_gr}, where a relevant operator of unknown strength is present and the boundary slope becomes a parameter that must be reconstructed from the data --- precisely the flat direction discussed there.
The UV structure of~\eqref{eq:Vnorm} otherwise mirrors the two-function parametrization~\eqref{eq:Vmodel_encoding}, so that both inverse problems are treated within the same family of encodings.

The area functional~\eqref{eq:Areg} is modified to use $f_{\text{NN}}(z)$ in place of the known blackening factor.
The remainder integral $\int_{z_{\text{mid}}}^{z_h} dz/(z^3\sqrt{f})$ requires special treatment because $f_{\text{NN}}(z_h) = 0$ produces a square-root singularity.
We use the change of variables $z(t) = z_h - (z_h - z_{\text{mid}})(1-t)^2$, whose Jacobian $dz/dt = 2(z_h - z_{\text{mid}})(1-t)$ analytically cancels the $1/\sqrt{z_h - z}$ divergence, yielding a bounded integrand that can be evaluated with the trapezoidal rule on a uniform $t$-grid of 300~points.

The alternating optimization proceeds as follows.
On even steps, the V-model is updated to minimize the data loss
\begin{equation}
  \mathcal{L}_{\text{data}} = 100 \left( A_{\text{reg}}[z_{\text{NN}},\, f_{\text{NN}}] - S_{EE}^{\text{data}}(l) \right)^2,
\end{equation}
where the factor of 100 amplifies the data-fitting signal (following ref.~\cite{Filev:2025}).
On odd steps, the L-model is updated to minimize the physical loss
\begin{equation}
  \mathcal{L}_{\text{phys}} = A_{\text{reg}}[z_{\text{NN}},\, f_{\text{NN}}]\,,
\end{equation}
which drives the surface toward the area minimum for the current learned metric.
Both networks are initialized randomly with no prior knowledge of the blackening factor.

\subsection{Method validation: AdS-Schwarzschild}
\label{sec:inverse_h1}

To validate our alternating optimization framework, we first reconstruct the blackening factor $f(z)$ of the AdS-Schwarzschild metric ($h=1$) from $50$~points of $S_{EE}(l)$ data on $l \in [0.15,\, 0.75\,l_c]$. We train for 500\,000~epochs with asymmetric learning rates ($\eta_L = 10^{-4}$, $\eta_V = 5\times 10^{-4}$), taking ${\sim}\,55$~minutes on a single CPU core; to quantify reproducibility, the training is repeated for five independent random seeds, which control both the network initializations and the stochastic sampling sequences.

\begin{figure}[ht]
\centering
\includegraphics[width=\textwidth]{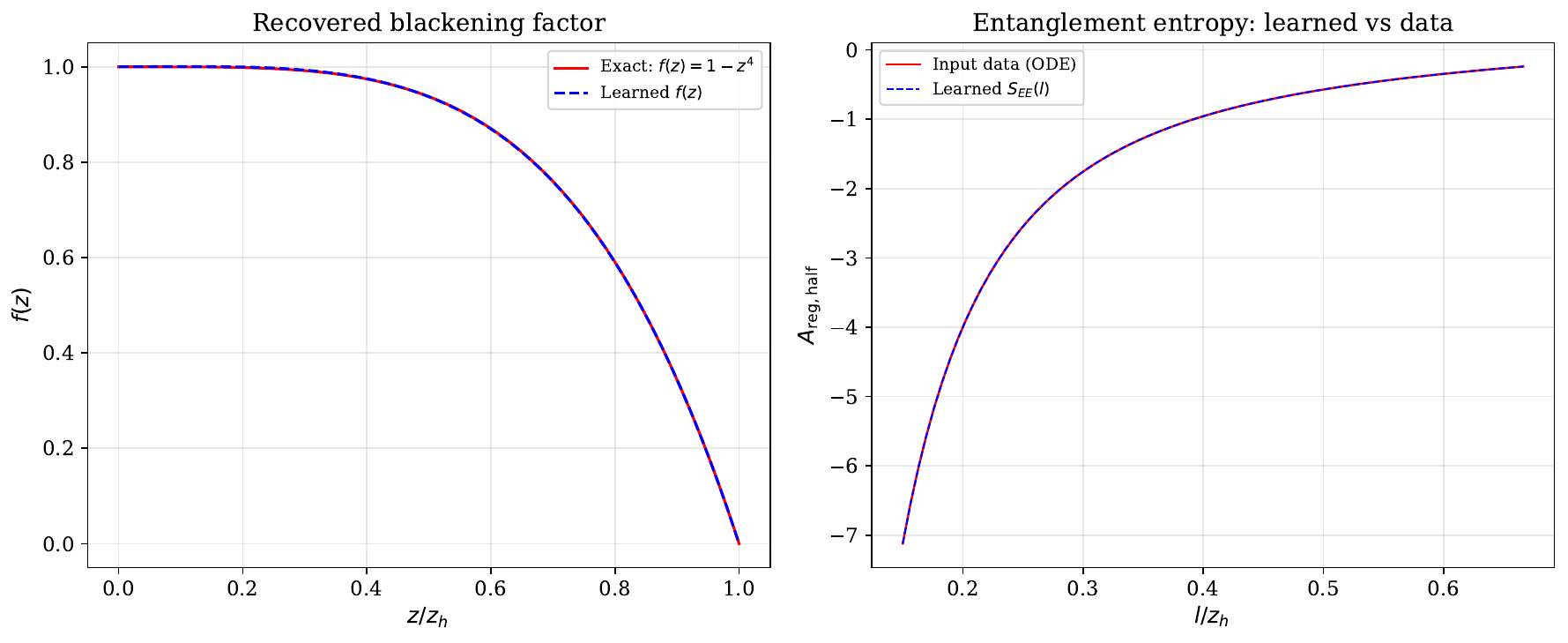}
\caption{Left: recovered blackening factor $f_{\text{NN}}(z)$ (dashed blue) compared with the exact $f(z) = 1 - z^4$ (solid red), for a representative one of the five independent runs.
Right: learned $S_{EE}(l)$ compared with the input data.
The ANN accurately reconstructs the metric from data alone.}
\label{fig:inverse}
\end{figure}

Across the five runs, the learned blackening factor matches the exact $f(z) = 1 - (z/z_h)^4$ with a maximum absolute error of $(1.9 \pm 0.9)\times10^{-3}$ over the entire bulk (worst seed $3.0\times10^{-3}$) and a mean absolute error of $(0.8 \pm 0.4)\times10^{-3}$; in relative terms, the maximum error for $z < 0.95\,z_h$ is $0.31\% \pm 0.15\%$.
The quoted maxima are robust rather than grid-point artifacts: the 95th percentile of the error profile lies within a few percent of the maximum.
The error reaches its floor within ${\sim}\,10^{5}$ epochs and shows no subsequent drift over the remaining $4\times10^{5}$ epochs; the residual floor is set by the finite data window, the interpolation of the input data, and the discretization of the integrals.

Since the $h=1$ case admits an exact semi-analytical inversion~\cite{Bilson:2008,Bilson:2010ff}, this validates the ANN variational method against a known benchmark without explicitly solving the Euler--Lagrange equations, reproducibly across independent training runs. We now turn to the genuinely degenerate problem of recovering multiple metric functions.

\subsection{Metric reconstruction at finite density: Gubser--Rocha}
\label{sec:inverse_gr}

The AdS-Schwarzschild background of the previous section has a single metric function $f(z)$, with $h(z) = 1$.
Many holographic models of physical interest, however, involve a non-trivial spatial warp factor $h(z) \neq 1$.
A prominent example is the Gubser--Rocha (GR) model~\cite{Gubser:2009qt}, an Einstein--Maxwell--Dilaton solution that describes a strongly coupled system at finite temperature and finite charge density.
Its dual field theory exhibits linear-in-$T$ resistivity and vanishing zero-temperature entropy --- hallmarks of strange metallic behavior in cuprate superconductors --- making it one of the most widely studied holographic models of condensed matter physics.
The reconstruction of the Gubser--Rocha bulk metric from entanglement entropy data has been addressed previously using neural ODEs~\cite{Ahn:2024}, while Transformer-based reconstruction of a blackening factor from entanglement data has been demonstrated in AdS$_3$~\cite{Kim:2025}; here we apply the variational approach.

To enable direct comparison with ref.~\cite{Ahn:2024}, we work with the AdS$_4$ ($d=3$ boundary) version of the GR model with vanishing momentum dissipation ($\beta = 0$ in the notation of ref.~\cite{Ahn:2024}), for which the metric functions admit closed-form expressions.
The metric takes the form
\begin{equation}\label{eq:metric_general}
  ds^2 = \frac{1}{z^2}\left[-f(z)\,dt^2 + \frac{dz^2}{f(z)} + h(z)\left(dx^2 + dy^2\right)\right],
\end{equation}
where $z \in [0, 1]$ with the AdS boundary at $z = 0$ and the horizon at $z_h = 1$, and~\cite{Ahn:2023}
\begin{equation}\label{eq:gr_solution}
  f(z) = (1-z)\,U(z)\,, \qquad
  h(z) = (1+Qz)^{3/2}\,,
\end{equation}
with
\begin{equation}\label{eq:U_def}
  U(z) = \frac{1 + (1+3Q)\,z + (1+3Q+3Q^2)\,z^2}{(1+Qz)^{3/2}}\,.
\end{equation}
Here $Q \geq 0$ is the charge parameter; $Q = 0$ recovers $f(z) = 1 - z^3$ and $h(z) = 1$ (AdS$_4$-Schwarzschild).
Note that $h(0) = 1$ and $h'(0) = \tfrac{3}{2}\,Q$; the non-vanishing boundary derivative reflects the fact that the coordinate $z$ is not the Fefferman--Graham radial variable, but is chosen to yield the closed-form expressions~\eqref{eq:gr_solution}.

The thermodynamic quantities are
\begin{equation}\label{eq:thermo_gr}
  T = \frac{3\sqrt{1+Q}}{4\pi}\,, \qquad
  s = \frac{(1+Q)^{3/2}}{4\,G_N}\,.
\end{equation}

The RT area functional for a strip of width~$l$ in the $x$-direction, with $\Omega = \int dy$ the transverse length, is
\begin{equation}\label{eq:area_gr}
  A = \Omega \int_{-l/2}^{l/2} dx\;\frac{\sqrt{h(z)}}{z^2}\,\sqrt{h(z) + \frac{z'^2}{f(z)}}\,,
\end{equation}
which reduces to the $d=3$ version of~\eqref{eq:area} when $h = 1$.
The first integral of the Euler--Lagrange equation gives a conserved Hamiltonian
\begin{equation}\label{eq:hamiltonian_gr}
  \mathcal{H} = \frac{h(z)^{3/2}}{z^2\,\sqrt{h(z) + z'^2/f(z)}} = \frac{h(z_*)}{z_*^2}\,,
\end{equation}
where $z_*$ is the turning point ($z' = 0$), which yields the parametric integrals
\begin{align}
  \frac{l}{2} &= \int_0^{z_*}
    \frac{dz}{\sqrt{f\,h}\;\sqrt{h^2 z_*^4/(z^4\,h_*^2) - 1}}\,,
  \label{eq:l_gr} \\[4pt]
  A_{\text{reg}} &= \Omega\!\int_0^{z_*}
    \frac{\sqrt{h}}{z^2\sqrt{f}}\left[\frac{1}{\sqrt{1-z^4 h_*^2/(z_*^4 h^2)}} - 1\right]dz
    - \Omega\!\int_{z_*}^{1}\frac{\sqrt{h}}{z^2\sqrt{f}}\,dz\,,
  \label{eq:Areg_gr}
\end{align}
where $h_* \equiv h(z_*)$.
The entanglement entropy $S_{EE}(l)$ depends on both $f(z)$ and $h(z)$ through this functional.
An important identity, due to ref.~\cite{Ahn:2024}, relates the derivative of the entropy directly to $h$ at the turning point:
\begin{equation}\label{eq:dSdl_gr}
  \frac{dS_{EE}}{dl} = \frac{\Omega}{4G_N}\,\frac{h(z_*)}{z_*^2}\,.
\end{equation}

The parametric integrals~\eqref{eq:l_gr}--\eqref{eq:Areg_gr} are used to generate the training data $S_{EE}(l)$ from the exact metric~\eqref{eq:gr_solution}.
The inverse problem, however, does \emph{not} use the turning-point parametrization.
Instead, we apply the same variational approach as in the AdS-Schwarzschild case:
the L-model learns the RT surface $z(x)$ directly, and the regularized area is computed using the hybrid $x/z$ integration scheme of section~\ref{sec:reg}, generalized to include $h(z)$.
Splitting at $x_s = l/5$ as before, the regularized half-area becomes
\begin{align}\label{eq:Areg_hybrid_gr}
  A_{\text{reg,half}} &= \int_0^{x_s}\! \frac{\sqrt{h}}{z^2}\sqrt{h + \frac{z'^2}{f}}\;dx
    + \int_\epsilon^{z_{\text{mid}}} \frac{\sqrt{h}}{z^2}\left[\sqrt{h\,x'^2 + \frac{1}{f}} - \frac{1}{\sqrt{f}}\right] dz
    \notag\\[4pt]
    &\quad - \int_{z_{\text{mid}}}^{z_h} \frac{\sqrt{h}}{z^2\sqrt{f}}\;dz\,,
\end{align}
where $x' = 1/z'$ and $z_{\text{mid}} = z(x_s)$.
The first term is the connected area in $x$-parametrization (interior), the second is the connected-minus-disconnected area in $z$-parametrization (boundary), and the third subtracts the remaining disconnected area from $z_{\text{mid}}$ to the horizon.
This is the key difference from the neural ODE approach of ref.~\cite{Ahn:2024}, which works with the turning-point integrals~\eqref{eq:l_gr}--\eqref{eq:Areg_gr}.

\subsubsection{The metric degeneracy}

A single function $S_{EE}(l)$ cannot uniquely determine two independent functions of~$z$.
To see why, we introduce following Bilson~\cite{Bilson:2010ff} the coordinate $r = z/\sqrt{h(z)}$, in which the metric~\eqref{eq:metric_general} takes the form
\begin{equation}\label{eq:metric_r}
  ds^2 = \frac{1}{r^2}\left[-g(r)\,e^{-\chi(r)}\,dt^2 + \frac{dr^2}{g(r)} + dx^2 + dy^2\right],
\end{equation}
with
\begin{equation}\label{eq:g_chi_def}
  g(r) = \alpha(z)^2\,f(z)\,,\qquad
  \chi(r) = \log\!\left[\alpha(z)^2\,h(z)\right],\qquad
  \alpha(z) \equiv 1 - \frac{z\,h'(z)}{2\,h(z)}\,.
\end{equation}
The spatial part of~\eqref{eq:metric_r} involves only $g(r)$; the function $\chi(r)$ appears exclusively in $g_{tt}$ and is invisible to any static minimal surface.
Bilson's inversion formula~\cite{Bilson:2008,Bilson:2010ff} recovers $g(r)$ from $S_{EE}(l)$ analytically:
\begin{equation}\label{eq:bilson_inversion}
  \frac{1}{\sqrt{g(r)}} = \frac{2}{\pi}\,\frac{1}{r^2}\,\frac{d}{dr}\int_0^r \frac{r_*^3}{\sqrt{r^4 - r_*^4}}\;\ell(r_*)\,dr_*\,,
\end{equation}
where $r_*$ is determined by $dS_{EE}/dl = \Omega/(4G_N\,r_*^2)$.
Thus $S_{EE}(l)$ uniquely determines $g(r)$ but provides no information about~$\chi(r)$.

For a fixed $g(r)$, any smooth $h(z) > 0$ with $\alpha > 0$ yields a valid metric via $f(z) = g(r(z))/\alpha(z)^2$ that produces the same $S_{EE}(l)$ (see appendix~\ref{app:degeneracy} for the detailed proof).
The thermal entropy and temperature impose only isolated point conditions at the horizon --- $h(z_h)$ and $h'(z_h)$ respectively --- which, together with the boundary value $h(0) = 1$, cannot determine the function in the bulk.
The boundary derivative $a = f'(0) = h'(0)$ is a free parameter not constrained by the data.
This degeneracy is \emph{exact} and has immediate consequences for any attempt to reconstruct both metric functions from entanglement data alone.

\begin{figure}[ht]
\centering
\includegraphics[width=0.75\textwidth]{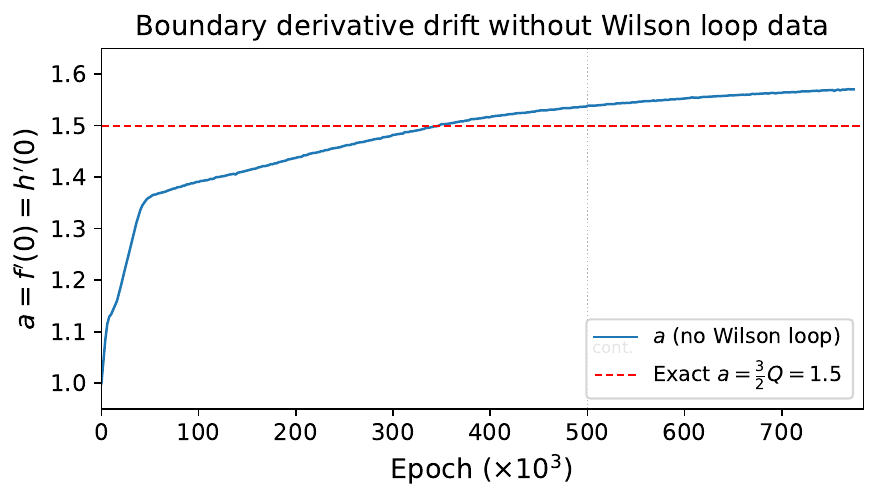}
\caption{Evolution of the boundary derivative $a = f'(0) = h'(0)$ during inverse training with $S_{EE}(l)$ and a thermal entropy penalty only (no Wilson loop data).
The dashed red line marks the exact value $a = \tfrac{3}{2}Q = 1.5$.
The parameter drifts past the exact value and continues to increase, explicitly revealing the flat direction in the loss landscape predicted by the metric degeneracy.}
\label{fig:a_drift}
\end{figure}

To demonstrate this explicitly, we attempt the inverse problem for the Gubser--Rocha metric using only $S_{EE}(l)$ together with a thermal entropy penalty $\lambda_s\bigl(h_{\text{NN}}(z_h) - s_{\text{target}}\bigr)^2$ that enforces the correct horizon value $h(z_h) = (1+Q)^{3/2}$.
The V-model parametrizes $f(z)$ and $h(z)$ with a shared trainable boundary derivative $a = f'(0) = h'(0)$ (initialized at $a = 1$, exact value $a = \tfrac{3}{2}Q = 1.5$).
Figure~\ref{fig:a_drift} shows the evolution of~$a$ over $7.7 \times 10^5$~epochs of training.
The parameter passes through the exact value around epoch~$350\,000$ but does not stabilize: it continues to drift upward, reaching $a \approx 1.57$ (a~$4.7\%$ error) with no sign of convergence.
Throughout this drift the $S_{EE}$ data-fitting loss remains small, confirming that the flat direction in the loss landscape corresponds precisely to the mathematical degeneracy identified above.

We note that Ahn et al.~\cite{Ahn:2024} reported a successful simultaneous reconstruction of $f(z)$ and $h(z)$ using neural ODEs with a thermodynamic penalty. Since the mathematical degeneracy proven above is exact, any method that appears to reconstruct both metric functions from $S_{EE}(l)$ alone must incorporate additional information beyond the entanglement data --- whether through explicit physical constraints, implicit biases of the network architecture, or regularization from the optimization dynamics. Our extended training run in figure~\ref{fig:a_drift} uses the variational area functional with minimal implicit bias, and the unconstrained parameter~$a$ drifts indefinitely, explicitly revealing the flat direction caused by this fundamental degeneracy. To achieve a mathematically stable, model-independent reconstruction, we must supplement the entanglement entropy with additional physical data.

\subsubsection{Breaking the degeneracy with the Wilson loop}

Breaking the degeneracy requires data sensitive to $\chi(r)$, i.e., to the timelike metric component.
Static minimal surfaces of any shape probe only the spatial geometry.
It is natural to ask whether genuinely Lorentzian entanglement observables could supply the missing information instead.
For boundary regions lying on a constant-time slice they cannot: in a static background, the HRT extremal surface~\cite{Hubeny:2007xt} lies on the corresponding constant-time bulk slice by time-reflection symmetry and reduces to the RT surface, so the degeneracy persists for \emph{all} equal-time entanglement data.
Entangling regions that do \emph{not} lie on a constant-time slice --- for example, strips anchored on boosted Cauchy slices --- are a genuine alternative: at finite temperature boundary boosts are not isometries, so the associated extremal surfaces extend in bulk time and couple to $g_{tt}$.
In this sense Wilson loop data is sufficient rather than fundamentally necessary to break the degeneracy.
Two considerations nevertheless favor the Wilson loop within the present framework.
First, Lorentzian extremal surfaces are saddle points of the area functional rather than minima --- variationally they are characterized by a maximin construction~\cite{Wall:2012uf} --- so they are not directly accessible to a method built on minimizing a loss function; a boosted-region variant of our approach would require a min-max training scheme.
Second, the temporal Wilson loop is a static equilibrium observable: it couples to the timelike metric while remaining a genuine minimization problem, fitting the variational framework without modification.
Time-dependent entanglement entropy following a quench would also probe $g_{tt}$, but takes the dual geometry outside the static class considered here.
What is needed within our static framework is thus an observable whose holographic dual extends along the time direction.

The holographic Wilson loop~\cite{Maldacena:1998im,Rey:1998ik} provides precisely such an observable.
Hashimoto \cite{Hashimoto:2020} derived inverse formulas recovering metric components from Wilson loop data; in the general case of two unknown metric functions, his method requires both temporal and spatial Wilson loops as independent inputs.
In the present work we use the temporal Wilson loop --- the screened test-charge potential --- in combination with entanglement entropy, and employ both a semi-analytical Bilson--Hashimoto inversion and a fully variational ANN implementation that does not require the turning-point reduction.
In the metric~\eqref{eq:metric_general}, the quark--antiquark potential is (see appendix~\ref{app:wilson} for the derivation):
\begin{equation}\label{eq:VL}
  V(L) = \frac{1}{2\pi\alpha'}\int_{-L/2}^{L/2}dx\;\frac{1}{z^2}\sqrt{f\,h + z'^2}\,,
\end{equation}
where the string worldsheet extends in time, coupling to $g_{tt}$ and hence to~$\chi$.
In the condensed matter context relevant to the Gubser--Rocha model, $V(L)$ computes the screened potential between two external test charges immersed in the strongly coupled medium~\cite{Rey:1998bq,Brandhuber:1998bs}.

The conserved Hamiltonian of the string yields the derivative relation (derived in appendix~\ref{app:wilson}):
\begin{equation}\label{eq:dVdL}
  \frac{dV_{\text{reg}}}{dL} = \frac{1}{2\pi\alpha'}\,\frac{\sqrt{f(\hat z_*)\,h(\hat z_*)}}{\hat z_*^2}\,,
\end{equation}
where $\hat z_*$ is the turning point of the string profile.
Combining entanglement entropy and Wilson loop data provides a complete reconstruction of the metric:
\begin{equation}\label{eq:sequential_reconstruction}
  S_{EE}(l) \;\xrightarrow{\text{Bilson}}\; g(r)\,,\qquad
  V(L) \;\xrightarrow{\text{Hashimoto}}\; \chi(r)\,,\qquad
  (g,\chi) \;\to\; (f,h) \text{ via \eqref{eq:g_chi_def}}\,.
\end{equation}
The first step uses Bilson's Abel inversion~\cite{Bilson:2008,Bilson:2010ff}; the second uses Hashimoto's Wilson loop inversion~\cite{Hashimoto:2020}, as described in the next section.

\subsubsection{Semi-analytical reconstruction: Bilson--Hashimoto method}
\label{sec:bilson_hashimoto}

We now present the complete semi-analytical reconstruction of the metric from $S_{EE}(l)$ and $V(L)$ boundary data, combining Bilson's entanglement inversion~\cite{Bilson:2008,Bilson:2010ff} with Hashimoto's Wilson loop inversion~\cite{Hashimoto:2020}.
The algorithm works entirely in the Bilson radial coordinate~$r$ and determines $g(r)$ and $\chi(r)$.

\medskip\noindent\textbf{Step 1: Bilson inversion $S_{EE} \to g(r)$.}
The Bilson coordinate $r_*$ of the RT turning point is extracted from the entanglement data via $dS_{EE}/dl = \Omega/(4G_N\,r_*^2)$~\cite{Ahn:2024}, giving $l(r_*)$ without knowledge of the metric.
The Bilson Abel integral~\eqref{eq:bilson_inversion} then yields $g(r)$.

\medskip\noindent\textbf{Step 2: Coordinate change.}
With $g(r)$ known, define a new radial coordinate
\begin{equation}\label{eq:eta_main}
  \eta(r) = \ln r + \int_0^r\!\left[\frac{1}{r'\sqrt{g(r')}} - \frac{1}{r'}\right]dr'\,,
\end{equation}
where the subtraction of the pure-AdS integrand $1/r'$ ensures convergence at the boundary.
The metric~\eqref{eq:metric_r} becomes
\begin{equation}\label{eq:metric_hash}
  ds^2 = -F(\eta)\,dt^2 + G(\eta)\!\left(dx^2+dy^2\right) + d\eta^2\,,
\end{equation}
with $G(\eta) = 1/r(\eta)^2$ (known from step~1) and $F(\eta)$ unknown.
This is precisely the metric form treated by Hashimoto~\cite{Hashimoto:2020}.

\medskip\noindent\textbf{Step 3: Hashimoto inversion $V(L) \to F(\eta)$.}
The Wilson loop derivative identity $h_0 = 2\pi\alpha'\,dV_{\rm reg}/dL = \sqrt{F_0 G_0}$ provides the parametric variable from boundary data.
Applying Hashimoto's Abel inversion~\cite{Hashimoto:2020} (see appendix~\ref{app:numerics} for implementation details), the structure function
\begin{equation}\label{eq:sigma_main}
  \sigma(H) = \frac{-1}{\pi}\;\frac{d}{dH}\int_H^{\infty}\frac{L(h_0)}{\sqrt{h_0^2 - H^2}}\,dh_0
\end{equation}
is determined, where $H = \sqrt{FG}$.
The relation $d\eta/dH = -\sigma(H)\sqrt{G(\eta)}$ is separable:
\begin{equation}\label{eq:match_main}
  \underbrace{\int_{-\infty}^{\eta}r(\eta')\,d\eta'}_{\Phi(\eta)}
  = \underbrace{\int_H^{\infty}\sigma(H')\,dH'}_{\Psi(H)}\,.
\end{equation}
Both integrals involve known functions.
Matching $\Phi(\eta) = \Psi(H)$ gives $H(\eta)$, hence
$F = H^2/G$ and
\begin{equation}\label{eq:chi_main}
  \chi(r) = \log\!\left[\frac{g(r)}{r^2\,F(\eta(r))}\right].
\end{equation}

\medskip\noindent\textbf{Step 4: Metric functions.}
From $g(r)$ and $\chi(r)$ in Bilson coordinates, the original metric functions are recovered via $h(z_*) = z_*^2/r_*^2$ and $f(z_*) = g(r_*)/\alpha^2$ with $\alpha = r_*\,e^{\chi(r_*)/2}/z_*$, where $r_*(z_*)$ is determined from the data (see appendix~\ref{app:numerics}).

\begin{figure}[ht]
\centering
\includegraphics[width=\textwidth]{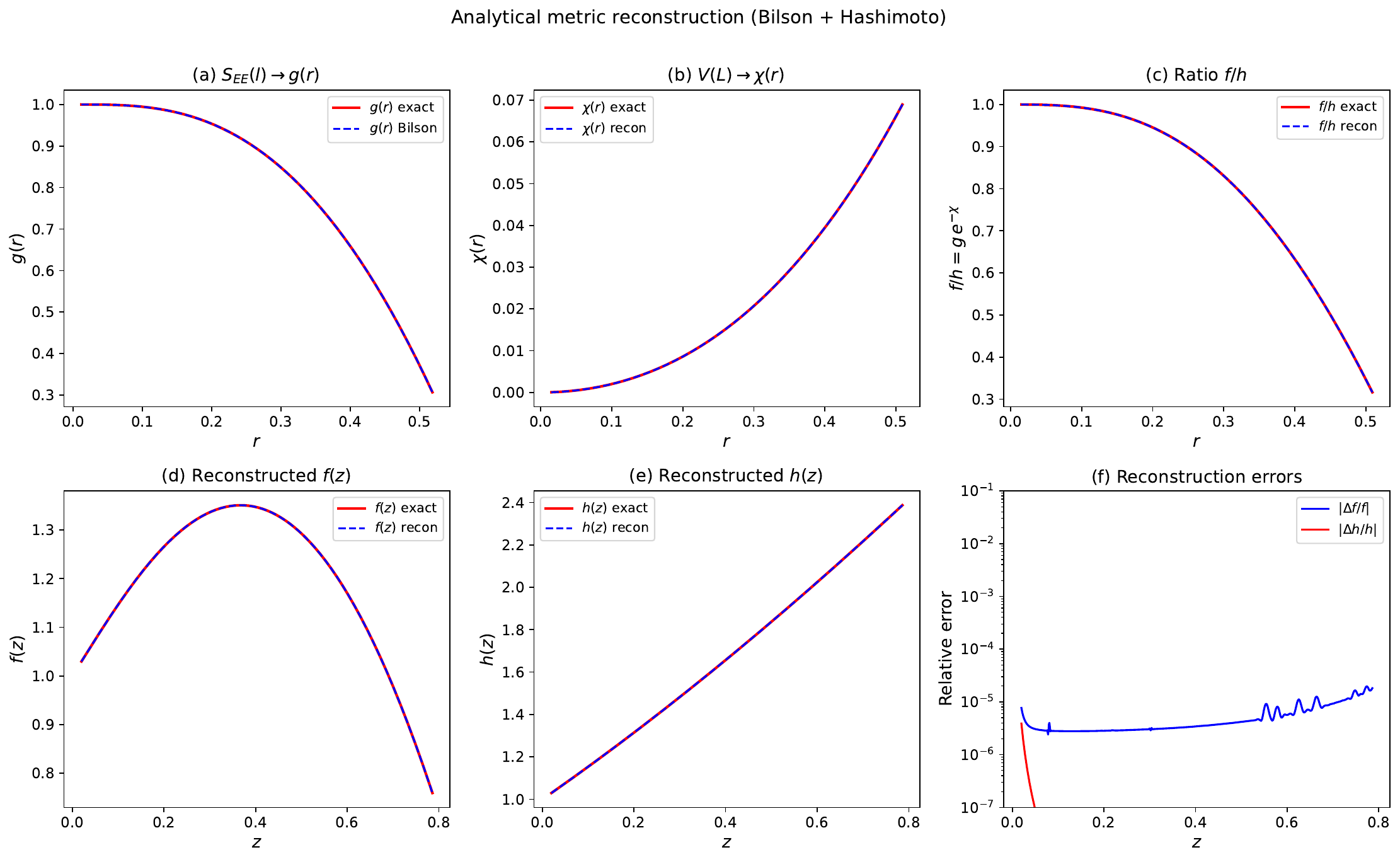}
\caption{Semi-analytical Bilson--Hashimoto metric reconstruction for the Gubser--Rocha model ($Q=1$).
Top row: (a)~$g(r)$ from Bilson inversion of $S_{EE}(l)$; (b)~$\chi(r)$ from Hashimoto inversion of $V(L)$; (c)~the ratio $f/h = g\,e^{-\chi}$.
Bottom row: (d)~reconstructed $f(z)$; (e)~reconstructed $h(z)$; (f)~relative errors.
Solid red: exact; dashed blue: reconstruction.
Both metric functions are recovered to better than $10^{-5}$ across the bulk.}
\label{fig:reconstruction}
\end{figure}

\medskip\noindent\textbf{Results.}
We demonstrate the reconstruction for the Gubser--Rocha metric with $Q = 1$, using 600 data points for $S_{EE}(l)$ and $V(L)$ generated from the exact metric.
The results are shown in figure~\ref{fig:reconstruction}.

The spatial metric $g(r)$ is recovered to a median relative error of $6 \times 10^{-9}$ from $S_{EE}$ alone.
Using the Wilson loop data, $\chi(r)$ is recovered to $3 \times 10^{-6}$, and the individual metric functions $f(z)$ and $h(z)$ to better than $2 \times 10^{-5}$.
This confirms that $S_{EE}$ and $V(L)$ together determine the complete metric, resolving the degeneracy with high precision.

\subsubsection{ANN approach with combined entanglement and Wilson loop data}

The Bilson--Hashimoto method of the previous section relies on closed-form derivative relations~\eqref{eq:dSdl_gr} and~\eqref{eq:dVdL} and the Abel-invertible structure of the turning-point integrals. For a more general numerical implementation that extends readily to other holographic observables (e.g., complexity or entanglement wedge cross-section) without requiring new semi-analytical derivations, we propose a variational approach using three neural networks with shared metric functions.

\begin{itemize}
  \item \textbf{L-model} (RT surface): a conditional network $z_L(x,l)$ parametrizing the RT minimal surface, trained by minimizing the area functional~\eqref{eq:area_gr}.
  \item \textbf{W-model} (Wilson loop string): a conditional network $z_W(x,L)$ parametrizing the string profile, trained by minimizing the Nambu--Goto action~\eqref{eq:VL}.
  \item \textbf{V-model} (metric): two sub-networks $f_{\text{NN}}(z)$ and $h_{\text{NN}}(z)$, shared between the L-model and W-model, encoding the bulk geometry.
\end{itemize}

The training proceeds via three-way alternating optimization:
\begin{enumerate}
  \item \emph{L-step}: update the L-model to minimize $A_{\text{reg}}[z_L,\,f_{\text{NN}},\,h_{\text{NN}}]$ for the current metric (drives the RT surface toward the area minimum).
  \item \emph{W-step}: update the W-model to minimize $V_{\text{reg}}[z_W,\,f_{\text{NN}},\,h_{\text{NN}}]$ for the current metric (drives the string toward the Nambu--Goto minimum).
  \item \emph{V-step}: update the V-model to minimize the combined data loss
  \begin{equation}\label{eq:loss_combined}
    \mathcal{L}_V = \frac{1}{N_l}\sum_{i=1}^{N_l}\left(A_{\text{reg}}(l_i) - S_{EE}^{\text{data}}(l_i)\right)^2
    + \frac{1}{N_L}\sum_{j=1}^{N_L}\left(V_{\text{reg}}(L_j) - V^{\text{data}}(L_j)\right)^2,
  \end{equation}
  which fits both the entanglement entropy and the Wilson loop data simultaneously.
\end{enumerate}
The entanglement entropy data constrains $g(r)$, while the Wilson loop data constrains the combination $g(r)\,e^{-\chi(r)}$; together they determine both $g$ and $\chi$, and hence both $f(z)$ and $h(z)$.

This approach requires Wilson loop data $V(L)$ as additional input.
In the holographic context, this is computed from the exact metric via~\eqref{eq:wl_Vreg}.
In the condensed matter interpretation of the Gubser--Rocha model, $V(L)$ corresponds to the screened potential between heavy external test charges in the strongly coupled medium~\cite{Rey:1998bq,Brandhuber:1998bs}, which is in principle accessible through impurity scattering experiments.

We implement this approach for the Gubser--Rocha metric with $Q=1$.
The L-model and W-model are conditional networks with two hidden layers of 32~neurons (1\,185~parameters each), using the same boundary condition encoding~\eqref{eq:bc_cond} with $d=3$.
(The reuse of the RT encoding for the string is justified in appendix~\ref{app:asymptotics}: the Nambu--Goto endpoint exponent is $1/3$ in \emph{any} boundary dimension, and coincides with the RT exponent $1/d$ precisely for $d = 3$.)
The V-model has two sub-networks $D_f$ and~$D_h$, each with two hidden layers of 20~neurons (963~parameters total including a shared trainable scalar~$a$).
Following the ansatz of ref.~\cite{Ahn:2024} and defining $\zeta \equiv z/z_h$, the metric functions are parametrized as
\begin{align}\label{eq:Vmodel_encoding}
  f_{\text{NN}}(z) &= 1 + a\,\zeta + \zeta^2\bigl[-(1+a) + D_f(\zeta) - D_f(1)\bigr], \notag\\
  h_{\text{NN}}(z) &= 1 + a\,\zeta + \zeta^2\,D_h(\zeta)\,,
\end{align}
where $a = f'(0)\,z_h = h'(0)\,z_h$ is initialized at $a = 1$ (exact value $a = \tfrac{3}{2}Q = 1.5$).
By construction $f_{\text{NN}}(0) = h_{\text{NN}}(0) = 1$, $f_{\text{NN}}(z_h) = 0$, and both functions share the same boundary derivative.
These conditions encode, respectively, membership in the asymptotically AdS class, the presence of a horizon, and a property of the UV expansion that is itself derived from the entanglement data in appendix~\ref{app:degeneracy}; the integer-power structure of~\eqref{eq:Vmodel_encoding} reflects the integer dimension of the dilaton operator of the dual theory (appendix~\ref{app:asymptotics}).
The value $h_{\text{NN}}(z_h)$ is left free and determined by the data.
The training data consists of 195~points of $S_{EE}(l)$ on $l \in [0.15,\, 0.73]$ (connected branch) and 150~points of $V(L)$ on $L \in [0.16,\, 0.54]$, both generated from the exact metric.

The training proceeds for 500\,000~epochs (128~minutes on a single CPU core) with learning rates $\eta_L = \eta_W = 10^{-4}$ and $\eta_V = 5\times 10^{-4}$, using a 4-step alternating cycle: V-step, L-step, V-step, W-step.
Crucially, the V-model loss contains \emph{only} the data-fitting terms for $S_{EE}$ and $V(L)$ --- no thermal entropy or temperature penalty is imposed.
The Wilson loop data alone is sufficient to break the degeneracy.

\begin{figure}[ht]
\centering
\includegraphics[width=\textwidth]{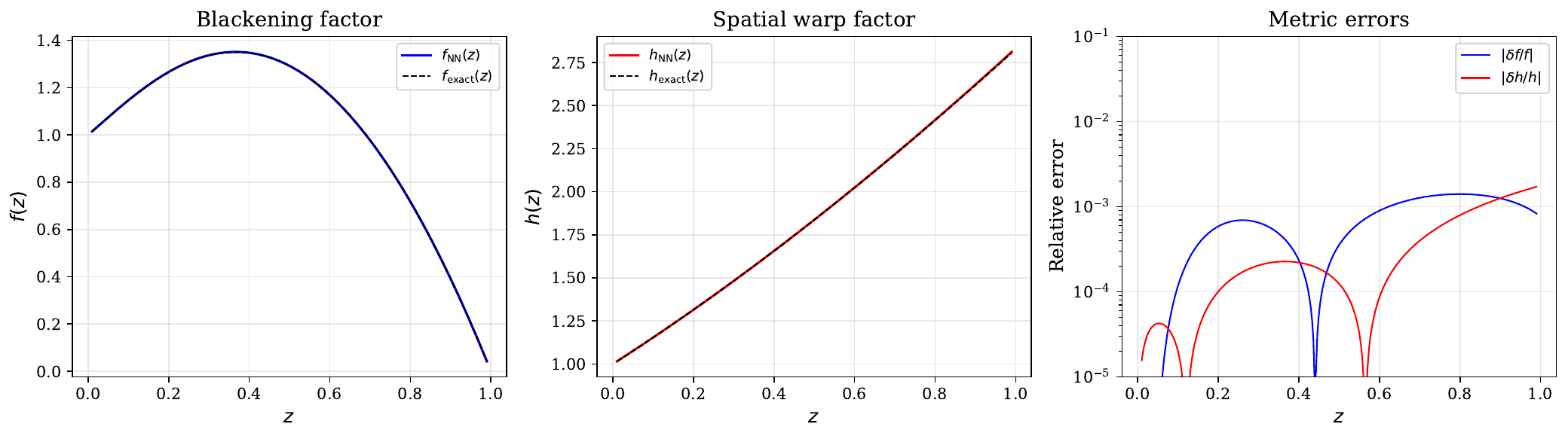}
\caption{Three-network metric reconstruction for the Gubser--Rocha model ($Q=1$) using combined $S_{EE}(l)$ and $V(L)$ data.
Left: recovered $f(z)$. Center: recovered $h(z)$. Right: relative errors.
Both metric functions are recovered to better than $0.2\%$ across the bulk, with no thermodynamic penalty.}
\label{fig:wl_metric}
\end{figure}

\begin{figure}[ht]
\centering
\includegraphics[width=\textwidth]{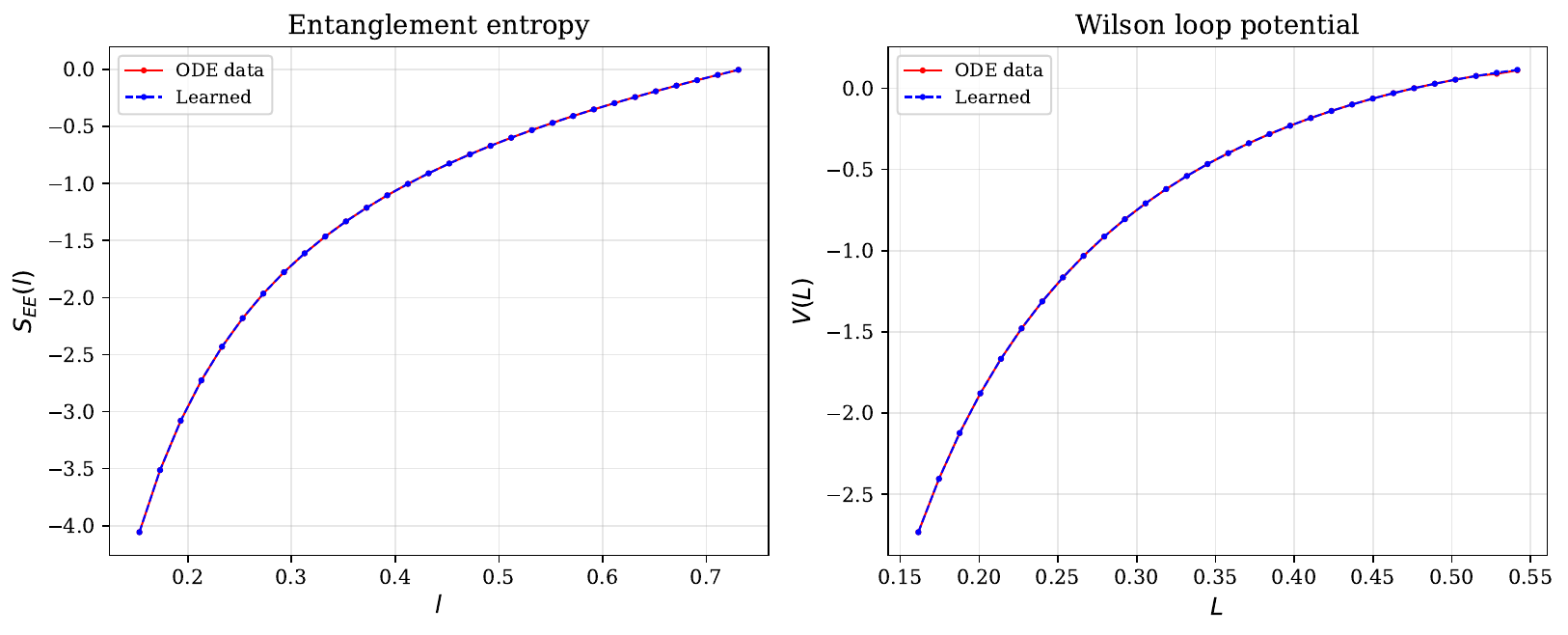}
\caption{Left: learned $S_{EE}(l)$ compared with input data.
Right: learned $V(L)$ compared with input data.
Both observables are reproduced to high accuracy by the jointly trained networks.}
\label{fig:wl_data_fit}
\end{figure}

\begin{figure}[ht]
\centering
\includegraphics[width=\textwidth]{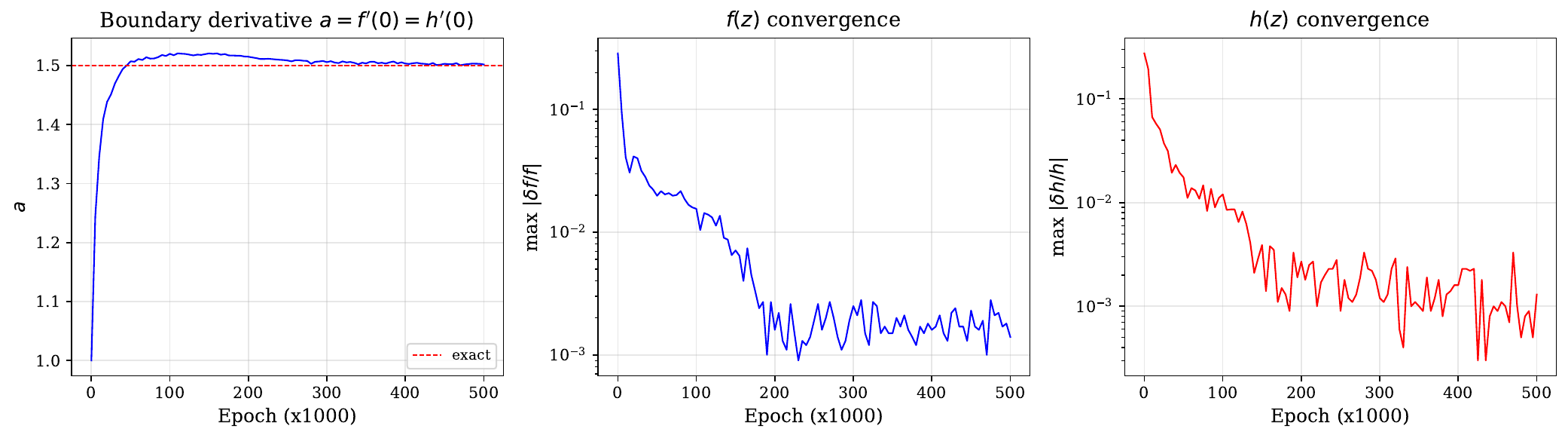}
\caption{Training convergence of the three-network approach.
Left: boundary derivative $a = f'(0) = h'(0)$ converges to the exact value $1.5$ within ${\sim}\,50\,000$~epochs and remains stable.
Center and right: maximum relative errors in $f(z)$ and $h(z)$ decrease to the sub-percent level.}
\label{fig:wl_convergence}
\end{figure}

The results are shown in figures~\ref{fig:wl_metric}--\ref{fig:wl_convergence}.
The learned blackening factor $f_{\text{NN}}(z)$ matches the exact $f(z)$ with a maximum relative error of $0.14\%$ (mean $0.07\%$) for $z < 0.99\,z_h$.
The spatial warp factor $h_{\text{NN}}(z)$ is recovered to $0.17\%$ maximum error (mean $0.04\%$).
The boundary derivative converges to $a = 1.5018$ (exact $1.5$), an error of $0.12\%$.
These figures are reproducible across independent training runs: ten random seeds at a shorter budget of $5\times10^4$ epochs all converge to the same solution with $a = 1.501 \pm 0.007$, and four seeds trained to the full $5\times10^5$ epochs give $\max|\Delta f/f| = 0.17\% \pm 0.06\%$, $\max|\Delta h/h| = 0.16\% \pm 0.14\%$, and $a = 1.5026 \pm 0.0022$ (section~\ref{sec:noise}).

\begin{figure}[ht]
\centering
\includegraphics[width=\textwidth]{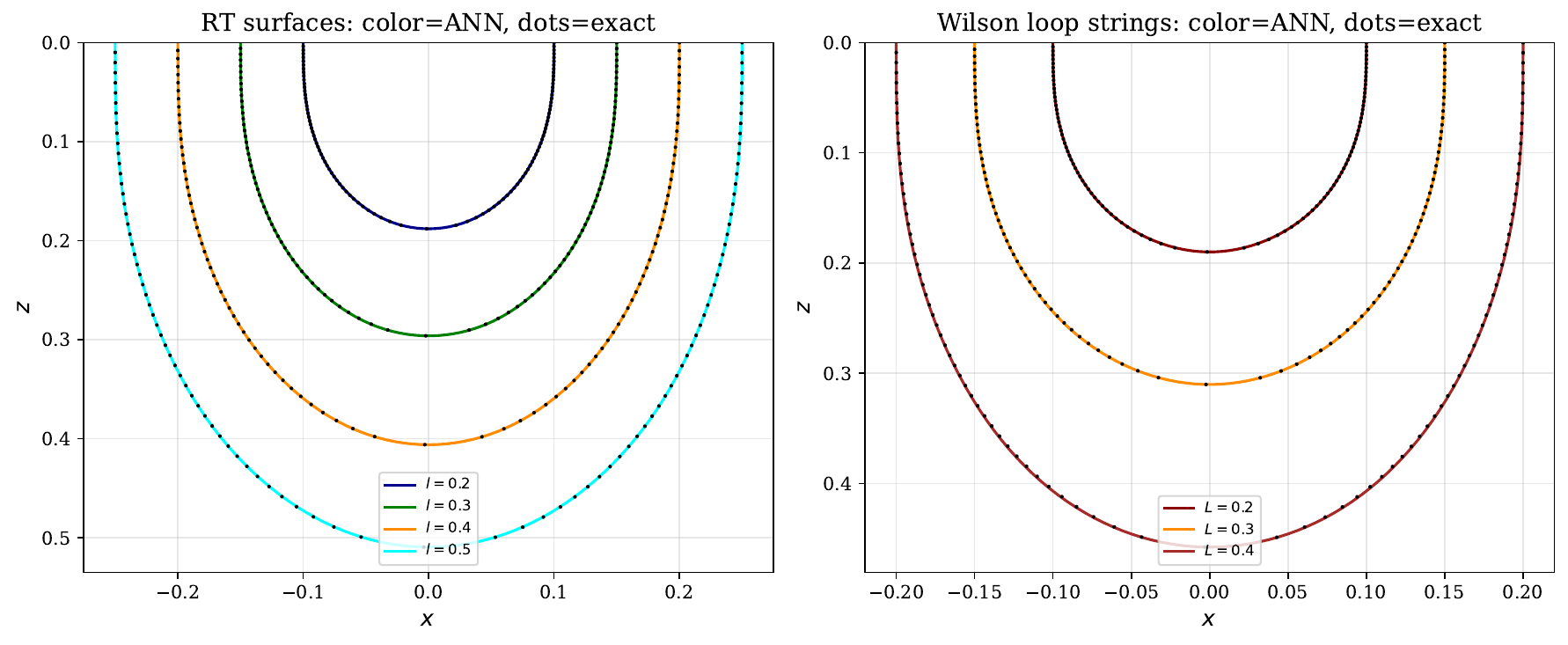}
\caption{Left: RT surface profiles $z_L(x)$ from the L-model (colored) vs exact (dots) for four strip widths.
Right: Wilson loop string profiles $z_W(x)$ from the W-model (colored) vs exact (dots) for three quark separations.
The ANN profiles are visually indistinguishable from the exact solutions.}
\label{fig:wl_profiles}
\end{figure}

The three-network approach recovers both metric functions to sub-$0.2\%$ accuracy ($0.14\%$ on~$f$, $0.17\%$ on~$h$), while the semi-analytical Bilson--Hashimoto method of section~\ref{sec:bilson_hashimoto} achieves ${\sim}\,10^{-5}$.
Both methods confirm that the variational framework correctly exploits the complementary information in $S_{EE}$ and $V(L)$.
Unlike the semi-analytical Bilson--Hashimoto method (which we found can exhibit severe numerical fragility such as roundoff errors and integration non-convergence when implemented computationally), the ANN approach does not require computing derivatives of the data or performing Abel inversions; it learns the metric directly from the raw observables.
The absence of any thermodynamic penalty demonstrates that the Wilson loop data --- or equivalently, the screened test-charge potential in the condensed matter context --- alone resolves the $f$--$h$ degeneracy, consistent with the theoretical argument of section~\ref{sec:inverse_gr}.

\subsection{Robustness of the reconstruction}
\label{sec:noise}

The reconstructions presented above correspond to individual training runs.
In this section we quantify their robustness with respect to the stochastic elements of the method --- the random initialization of the networks and the sampling sequences during training --- and with respect to noise in the boundary data.
Both observables enter every reconstruction: as established in section~\ref{sec:inverse_gr}, entanglement entropy alone leaves an exact flat direction, so the joint $S_{EE}$--$V(L)$ data set is the setting in which the robustness of the method is meaningfully assessed.

\begin{figure}[ht]
\centering
\includegraphics[width=0.8\textwidth]{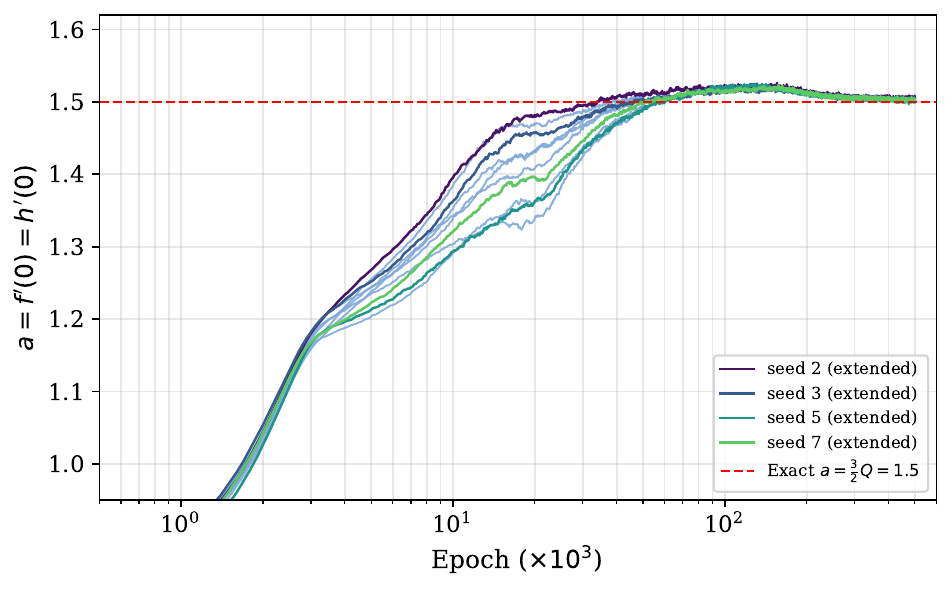}
\caption{Training trajectories of the boundary derivative $a = f'(0) = h'(0)$ for the three-network reconstruction with combined $S_{EE}(l)$ and $V(L)$ data: ten independent random seeds trained for $5\times10^4$ epochs (thin light curves), four of which are extended to $5\times10^5$ epochs (colored curves).
Every run converges to the exact value $a = \tfrac{3}{2}Q = 1.5$ (dashed red line) and remains locked there --- in contrast with the unbounded drift of figure~\ref{fig:a_drift} when the Wilson loop data is removed.}
\label{fig:a_seeds}
\end{figure}

\paragraph{Seed dependence.}
We repeated the three-network reconstruction for ten independent random seeds, which control the initialization of all three networks as well as the stochastic sampling of strip widths and quark separations during training.
At a fixed budget of $5\times10^4$ epochs all ten runs converge to the same solution, with $a = 1.501 \pm 0.007$ and maximum relative errors of $2.4\% \pm 0.2\%$ on~$f$ and $2.3\% \pm 0.2\%$ on~$h$ (mean $\pm$ standard deviation across seeds).
Four of the seeds were trained to the full $5\times10^5$ epochs; the results, collected in table~\ref{tab:seeds}, show that the sub-$0.2\%$ accuracy reported in section~\ref{sec:inverse_gr} is attained reproducibly.
Figure~\ref{fig:a_seeds} shows the training trajectories of the boundary derivative for all runs: every seed locks onto the exact value $a = \tfrac{3}{2}Q = 1.5$ and remains there, in sharp contrast with the unbounded drift of figure~\ref{fig:a_drift} observed when the Wilson loop data is removed.

\begin{table}[ht]
\centering
\begin{tabular}{lcccc}
\toprule
Seed & $a$ (exact $1.5$) & Max $|\Delta f/f|$ & Max $|\Delta h/h|$ & $h(z_h)$ (exact $2.828$) \\
\midrule
2 & 1.5049 & 0.10\% & 0.35\% & 2.839 \\
3 & 1.5035 & 0.22\% & 0.05\% & 2.830 \\
5 & 1.4998 & 0.22\% & 0.07\% & 2.827 \\
7 & 1.5021 & 0.15\% & 0.18\% & 2.834 \\
\midrule
mean $\pm$ std & $1.5026 \pm 0.0022$ & $0.17\% \pm 0.06\%$ & $0.16\% \pm 0.14\%$ & $2.832 \pm 0.005$ \\
\bottomrule
\end{tabular}
\caption{Seed dependence of the three-network reconstruction at $5\times10^5$ epochs.
All seeds recover both metric functions at the sub-$0.2\%$ level, with the boundary derivative and horizon warp factor statistically consistent with their exact values.}
\label{tab:seeds}
\end{table}

\paragraph{Noise robustness.}
In any practical application the boundary data will carry numerical, experimental, or lattice uncertainties.
We model this by corrupting \emph{both} observables with independent multiplicative Gaussian noise,
\begin{equation}
  S_{EE}^{\text{noisy}}(l_i) = S_{EE}(l_i)\,\bigl(1 + \sigma\,\xi_i\bigr), \qquad
  V^{\text{noisy}}(L_j) = V(L_j)\,\bigl(1 + \sigma\,\eta_j\bigr), \qquad
  \xi_i,\,\eta_j \sim \mathcal{N}(0,1)\,,
\end{equation}
while the exact metric is used only to evaluate the reconstruction error.
For each noise level $\sigma \in \{1\%,\, 5\%\}$ we perform five independent reconstructions, each with its own noise realization and random seed, at a fixed budget of $3\times10^5$ epochs; the ensemble errors plateau by ${\sim}2.5\times10^5$ epochs, so the results below reflect converged training.
The noiseless ensemble at the same budget provides the intrinsic accuracy floor of the method.

\begin{table}[ht]
\centering
\begin{tabular}{ccccc}
\toprule
Noise $\sigma$ & Realizations & Max $|\Delta f/f|$ & Max $|\Delta h/h|$ & $a$ (exact $1.5$) \\
\midrule
0    & 4 & $0.20\% \pm 0.02\%$ & $0.19\% \pm 0.13\%$ & $1.506 \pm 0.002$ \\
1\%  & 5 & $0.44\% \pm 0.20\%$ & $0.38\% \pm 0.16\%$ & $1.512 \pm 0.014$ \\
5\%  & 5 & $1.64\% \pm 0.61\%$ & $1.59\% \pm 0.57\%$ & $1.532 \pm 0.050$ \\
\bottomrule
\end{tabular}
\caption{Noise robustness of the three-network reconstruction (mean $\pm$ standard deviation across realizations, all at a fixed budget of $3\times10^5$ epochs).
Both the error and its run-to-run spread grow monotonically with the noise amplitude.
The small upward bias of $a$ in the noiseless row is a residual finite-budget transient, which relaxes to $a = 1.5026 \pm 0.0022$ at $5\times10^5$ epochs (table~\ref{tab:seeds}).}
\label{tab:noise}
\end{table}

\begin{figure}[ht]
\centering
\includegraphics[width=\textwidth]{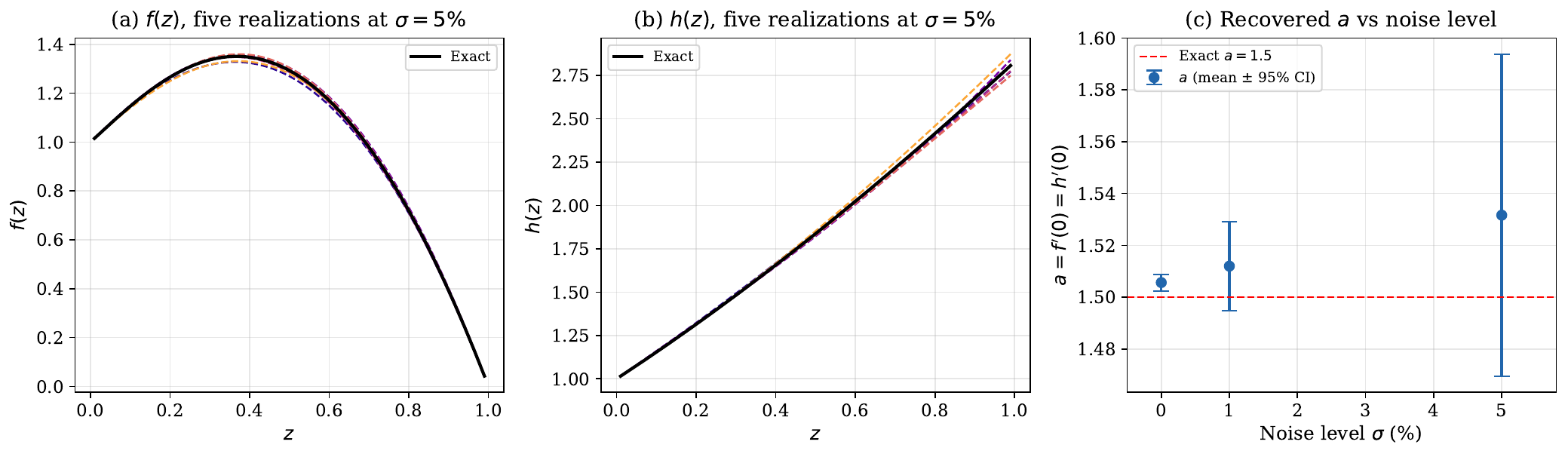}
\caption{Noise robustness of the three-network reconstruction with noise applied to both observables.
(a)~Recovered $f(z)$ for the five noise realizations at $\sigma = 5\%$ (dashed), overlaid on the exact function (solid).
(b)~The same for $h(z)$.
(c)~Recovered boundary derivative $a$ versus noise level, with $95\%$ confidence intervals over the realizations; the uncertainty grows with $\sigma$ while remaining consistent with the exact value $a = 1.5$ (dashed red line).}
\label{fig:noise_wl}
\end{figure}

Table~\ref{tab:noise} summarizes the results.
Both the reconstruction error and its run-to-run spread grow monotonically with the noise amplitude, from the intrinsic floor of ${\sim}0.2\%$ at $\sigma = 0$ --- set by the finite data window, the interpolation of the input data, and the discretization of the integrals --- to ${\sim}1.6\%$ at $\sigma = 5\%$.
The recovered boundary derivative remains statistically consistent with the exact value: the $95\%$ confidence intervals, $a = 1.512 \pm 0.017$ at $\sigma = 1\%$ and $a = 1.532 \pm 0.062$ at $\sigma = 5\%$, both contain $a = 1.5$, with an uncertainty that grows with~$\sigma$ (right panel of figure~\ref{fig:noise_wl}).
The left panels of figure~\ref{fig:noise_wl} overlay the five $\sigma = 5\%$ reconstructions on the exact metric functions: every realization tracks the exact curves at the percent level, and no realization exhibits erratic run-to-run behavior.

This smooth response to noise is not accidental: it is inherited from the well-posedness restored by the Wilson loop.
With both observables in the loss the inverse problem has no flat direction, and noise in the data perturbs the reconstruction smoothly instead of displacing the optimizer along unconstrained directions in the space of metrics.

\section{Comparison of ODE and ANN methods}
\label{sec:comparison}

We now summarize the accuracy and computational cost of the different approaches presented in this paper.

\subsection{Accuracy}

Table~\ref{tab:comparison} collects the accuracy of the three ANN methods against the ODE benchmark.

\begin{table}[ht]
\centering
\setlength{\tabcolsep}{5.5pt}
\begin{tabular}{lccccc}
\toprule
Method & Parameters & Epochs & Training time & $\delta A/A$ (mean) & $\delta A/A$ (max) \\
\midrule
Single-$l$ ANN     & 461 & 5\,000   & 37\,s / strip  & $3.0\times 10^{-4}$ & $5.6\times 10^{-4}$ \\
Conditional ANN    & 1\,021 & 120\,000 & 28\,min & $3.5\times 10^{-3}$ & $1.3\times 10^{-1}$ \\
Inverse (V-model)  & 481 & 500\,000 & 55\,min & $1.4\times 10^{-3}$ & $3.0\times 10^{-3}$ \\
\bottomrule
\end{tabular}
\caption{Comparison of ANN methods.
The single-$l$ ANN is trained per strip width; the conditional and inverse methods each produce a family of solutions in one training run.
The conditional max error is inflated by $A_{\text{reg}} \to 0$ at the phase transition.
The inverse-row errors are averaged over five independent training runs.
All computations are performed on a single CPU core.}
\label{tab:comparison}
\end{table}

For the forward problem, the single-$l$ ANN achieves ${\sim}\,0.03\%$ accuracy in 37~seconds per strip width.
The conditional network trades an order of magnitude in accuracy ($0.35\%$ mean error) for the ability to evaluate any~$l$ without retraining.
Its apparent maximum error of $13\%$ occurs at $l \approx l_c$, where $A_{\text{reg}} \to 0$; the absolute error at that point is still only ${\sim}\,3 \times 10^{-4}$.

For the inverse problem, the V-model recovers $f(z)$ with a maximum absolute error of $(1.9 \pm 0.9)\times10^{-3}$ across five independent runs ($0.31\% \pm 0.15\%$ maximum relative error for $z < 0.95\,z_h$) and reproduces $S_{EE}(l)$ to better than $0.5\%$.

\subsection{Convergence}

\begin{figure}[ht]
\centering
\includegraphics[width=\textwidth]{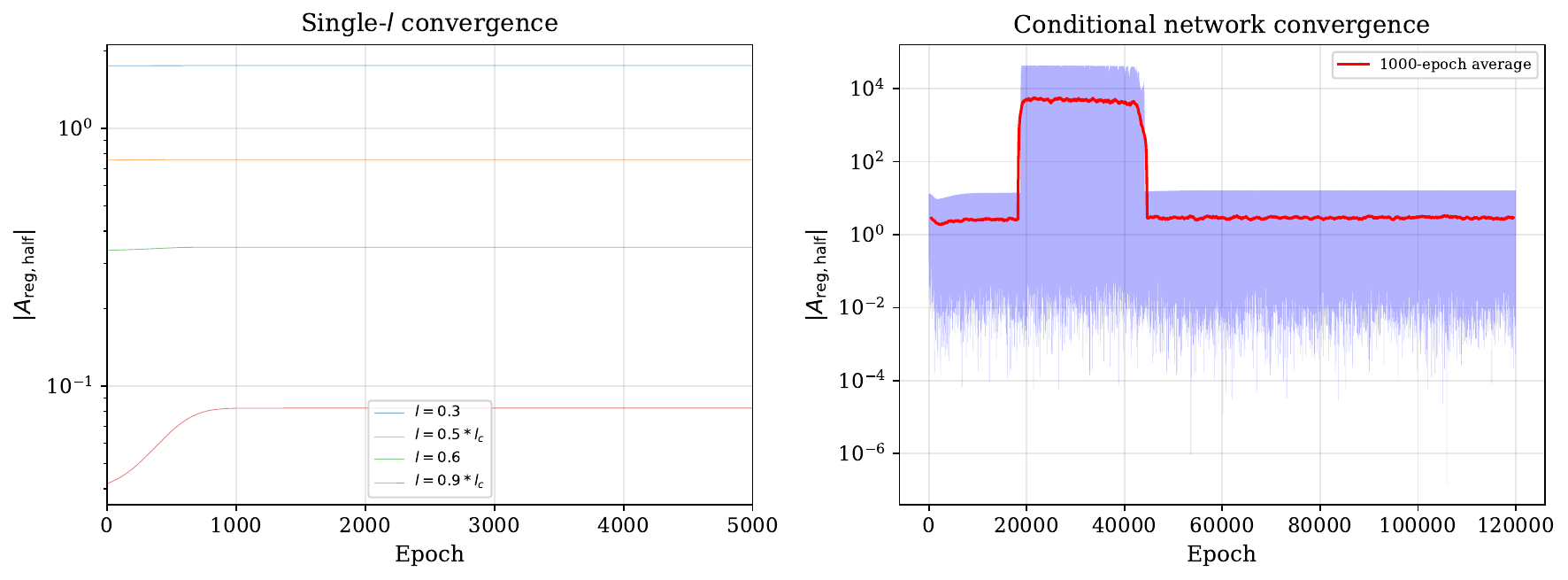}
\caption{Left: loss convergence for the four single-$l$ test cases.
Right: conditional network loss (light blue) and its 1000-epoch running average (red).}
\label{fig:convergence_comparison}
\end{figure}

Figure~\ref{fig:convergence_comparison} shows the training convergence for the single-$l$ and conditional approaches.
The single-$l$ loss stabilizes within ${\sim}\,1000$~epochs.
The conditional network converges more slowly, requiring ${\sim}\,50\,000$~epochs for the smoothed loss to plateau, reflecting the larger function space it must learn.

\begin{figure}[ht]
\centering
\includegraphics[width=\textwidth]{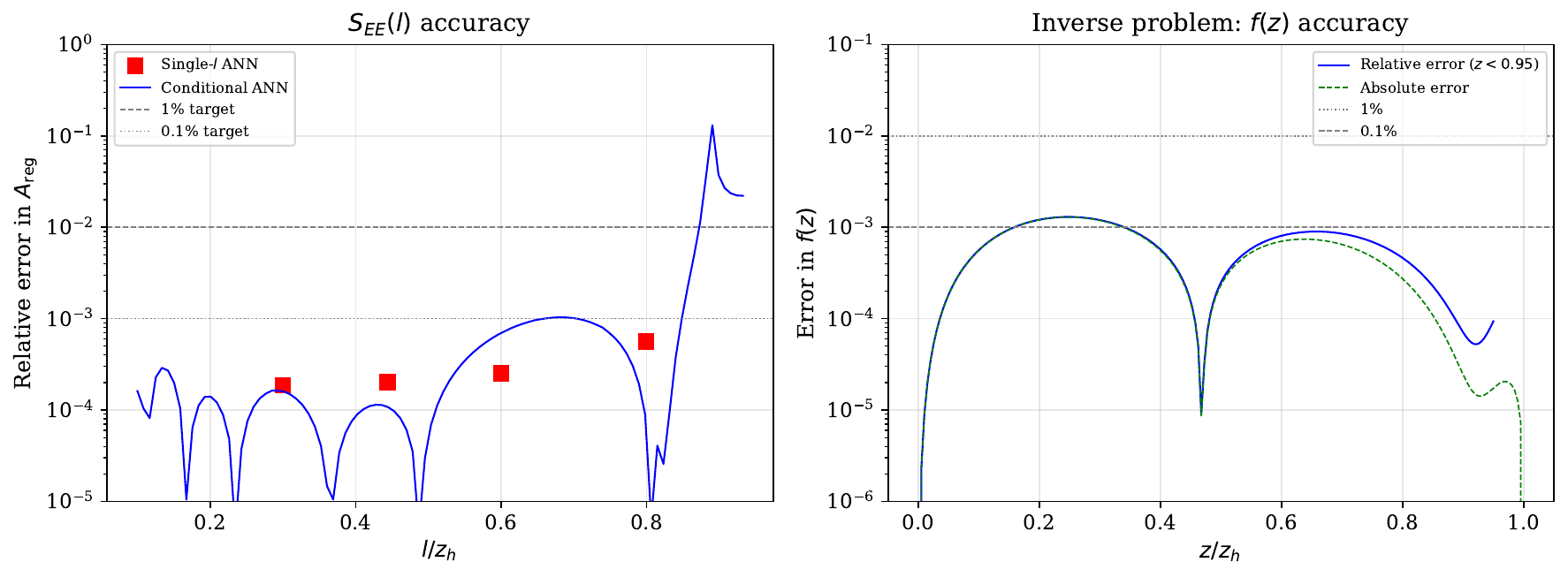}
\caption{Left: relative error in $A_{\text{reg}}$ vs strip width for the single-$l$ ANN (squares) and conditional ANN (curve).
Right: relative and absolute error in the recovered blackening factor $f(z)$ from the inverse problem.}
\label{fig:accuracy_summary}
\end{figure}

Figure~\ref{fig:accuracy_summary} provides a unified view of the accuracy across all methods.
The single-$l$ networks (red squares) achieve ${\sim}\,10^{-4}$ relative error uniformly.
The conditional network (blue curve) achieves ${\sim}\,10^{-3}$ over most of the $l$ range, with the expected spike at $l_c$.
The inverse problem (right panel) recovers $f(z)$ with absolute error below $3\times10^{-3}$ across the entire radial range; expressed as a relative error, the largest values occur near the horizon, where $f \to 0$.

\subsection{Computational cost}

The ODE benchmark (498~turning points, each requiring adaptive quadrature of two integrals) runs in ${\sim}\,10$~seconds.
For a single strip width, the ODE shooting method is faster than the ANN by a factor of ${\sim}\,200$.
However, the ANN approach offers two advantages over the ODE method:
\begin{enumerate}
  \item The conditional network provides the full $S_{EE}(l)$ curve as a differentiable function, enabling automatic differentiation with respect to~$l$ at negligible marginal cost.
  The ODE method would require a separate shooting computation for each~$l$.
  \item The inverse problem --- recovering $f(z)$ from $S_{EE}(l)$ data --- has no ODE counterpart.
  The ODE shooting method \emph{requires} knowing $f(z)$ a priori; it cannot reconstruct the metric from data.
  This is the decisive advantage of the ANN variational approach.
\end{enumerate}

All computations in this paper were performed in double precision (float64) on a single CPU core using PyTorch~\cite{pytorch} with the Adam optimizer~\cite{adam}.

\section{Conclusion}
\label{sec:conclusion}

We have introduced a general framework for holographic bulk reconstruction using artificial neural networks, where boundary data directly constrains the spacetime geometry through the variational minimization of area and action functionals.

The method accurately solves the forward problem for Ryu--Takayanagi minimal surfaces without explicitly deriving or solving the Euler--Lagrange equations, capturing the connected/disconnected phase transition. For the single-function inverse problem (AdS-Schwarzschild), the approach accurately recovers the blackening factor, validating the numerical framework against Bilson's semi-analytical inversion~\cite{Bilson:2008,Bilson:2010ff}.

The central physical result of this paper concerns the inverse problem for finite-density metrics with two unknown functions, $f(z)$ and $h(z)$, such as the Gubser--Rocha model. For the class of static, diagonal, translation-invariant backgrounds with strip entangling regions considered here, we proved that equal-time entanglement entropy determines only the spatial metric component. The timelike component remains entirely unconstrained, giving rise to an exact one-function degeneracy that cannot be lifted by thermodynamic point conditions. This intrinsic degeneracy provides a fundamental explanation for the instability observed during extended neural network training, which in the absence of additional data or implicit biases exhibits unbounded drift along flat directions in the parameter space.

To uniquely reconstruct the complete metric, one must probe the timelike geometry. We resolve the degeneracy by supplementing entanglement entropy with holographic Wilson loop data --- in the condensed matter interpretation of the Gubser--Rocha model, the screened potential between external test charges in the strongly coupled medium. The string worldsheet naturally extends in time, coupling to the timelike metric component and explicitly breaking the degeneracy.

We provided two complementary resolutions to the two-function inverse problem. First, a semi-analytical reconstruction combining Bilson's entanglement entropy inversion~\cite{Bilson:2008,Bilson:2010ff} with Hashimoto's Wilson loop inversion~\cite{Hashimoto:2020}, which sequentially determines $g(r)$ and $\chi(r)$ in the Bilson radial coordinate. Second, a general three-network variational approach that jointly minimizes the combined area and Nambu--Goto actions to recover $f(z)$ and $h(z)$ to better than $0.2\%$ accuracy. While the semi-analytical method requires closed-form derivative relations, the three-network model needs only the action functional. This makes the ANN variational approach naturally extensible: adding a new observable requires only introducing an additional network and loss term, rather than deriving a new semi-analytical inversion scheme.
We further quantified the robustness of the reconstruction: it is reproducible across independent random seeds, attaining the quoted accuracy with $a = 1.5026 \pm 0.0022$, and it degrades smoothly and monotonically under noise applied to both observables (section~\ref{sec:noise}).

For the class of static diagonal metrics studied here, this work illustrates a conceptual principle: the entanglement entropy of spatial subregions builds the spatial bulk geometry, while the Wilson loop --- through its coupling to the timelike metric --- completes the Lorentzian structure. It would be interesting to investigate whether this complementarity extends to more general metric classes and entangling region geometries, and whether Lorentzian entanglement data --- extremal surfaces anchored on boosted boundary regions, accessed through a min-max variational scheme --- could play the role of the Wilson loop, as discussed in section~\ref{sec:inverse_gr}.

\appendix

\section{Proof of the metric degeneracy}
\label{app:degeneracy}

We prove that $S_{EE}(l)$ determines only one combination of the two metric functions $f(z)$ and $h(z)$, leaving a family of physically distinct spacetimes --- one for every admissible choice of a free function --- that all produce identical entanglement entropy.

Before the details, let us describe in words what the proof does and what is held fixed.
Throughout, the quantity kept fixed is the boundary data itself: the entanglement entropy $S_{EE}(l)$ for all strip widths.
Because the surfaces that compute $S_{EE}$ lie on a slice of constant time, they sample only the spatial part of the metric; written in a suitable radial coordinate (Step~1), the entire spatial geometry is encoded in a single function $g(r)$, and the time-time component of the metric never enters.
Holding $g(r)$ fixed is therefore the same thing as holding all the entanglement data fixed.
The proof then deforms the metric without touching $g(r)$: one picks a new spatial warp factor $\tilde h$ essentially at will and compensates it with a new blackening factor $\tilde f$, constructed so that the two changes cancel exactly in the spatial geometry (Step~2).
The cancellation holds for every strip width at once --- this is not an approximate statement.
The deformed spacetime still looks the same at the boundary; matching the thermal entropy places its horizon at the same location, and matching the temperature fixes one further number --- point conditions that do not constrain the free function in the bulk (Step~3).
Step~4 exhibits explicit examples, and Step~5 shows, order by order in the near-boundary expansion, that each order of the data constrains one combination of the two new Taylor coefficients while always leaving the complementary combination free.
What \emph{does} change under the deformation is the time-time component of the metric --- and with it observables that extend in time, such as the Wilson loop --- which is the handle used in section~\ref{sec:bilson_hashimoto} to remove the degeneracy.

\paragraph{Step 1: Bilson coordinates.}
The coordinate transformation $r = z/\sqrt{h(z)}$ maps the metric~\eqref{eq:metric_general} to the Bilson form~\eqref{eq:metric_r}, with
\begin{equation}
  g(r) = \alpha(z)^2\,f(z)\,,\qquad
  \chi(r) = \log\!\left[\alpha(z)^2\,h(z)\right],\qquad
  \alpha(z) \equiv 1 - \frac{z\,h'(z)}{2\,h(z)}\,.
\end{equation}
The spatial part of the Bilson metric, $r^{-2}[dr^2/g + dx^2 + dy^2]$, depends only on~$g(r)$.
Since the RT area functional for any static entangling region involves only the spatial metric, $S_{EE}(l)$ is a functional of $g(r)$ alone.
The function $\chi(r)$ appears only in $g_{tt}$ and is invisible to all static minimal surfaces.
Note that $\alpha(0) = 1$ for any smooth $h$ with $h(0) = 1$, since the factor $z h'(z)/[2h(z)]$ vanishes at $z = 0$; combined with the asymptotically AdS boundary values $f(0) = h(0) = 1$, this gives $r \simeq z$ near the boundary and $g(0) = \alpha(0)^2 f(0) = 1$.

\paragraph{Step 2: Constructing the degenerate family.}
Bilson's inversion~\cite{Bilson:2008,Bilson:2010ff} uniquely determines $g(r)$ from $S_{EE}(l)$.
Now choose \emph{any} smooth function $\tilde h(z) > 0$ satisfying $\tilde h(0) = 1$ and $\tilde\alpha(z) > 0$, and define
\begin{equation}\label{eq:f_from_g}
  \tilde f(z) = \frac{g\!\bigl(\tilde r(z)\bigr)}{\tilde\alpha(z)^2}\,,\qquad
  \tilde r(z) = \frac{z}{\sqrt{\tilde h(z)}}\,,\qquad
  \tilde\alpha(z) = 1 - \frac{z\,\tilde h'(z)}{2\,\tilde h(z)}\,.
\end{equation}
By construction, the pair $(\tilde f, \tilde h)$ maps to the \emph{same} $g(r)$ as the original metric, and hence produces the same $S_{EE}(l)$.
However, $\tilde\chi(r) = \log[\tilde\alpha^2\,\tilde h] \neq \chi(r)$ in general, so the spacetime is physically different (different $g_{tt}$, different temperature, different causal structure).

\paragraph{Step 3: Boundary and horizon conditions.}
The boundary condition is automatically satisfied:
$\tilde f(0) = g(0)/\tilde\alpha(0)^2 = 1$, since $g(0) = 1$ (Step~1) and $\tilde\alpha(0) = 1$ (the argument of Step~1 applied to $\tilde h$, which is smooth with $\tilde h(0) = 1$).
At the horizon, the thermal entropy density $s = (1+Q)^{3/2}/(4G_N)$ fixes $\tilde h(z_h) = z_h^2/r_h^2$, providing one point constraint on~$\tilde h$; this in turn places the horizon of the deformed metric at $z_h$, since then $\tilde r(z_h) = r_h$ and $\tilde f(z_h) = g(r_h)/\tilde\alpha(z_h)^2 = 0$.
The temperature imposes one further point condition: differentiating~\eqref{eq:f_from_g} at the horizon, where $g(r_h) = 0$, gives $\tilde f'(z_h) = g'(r_h)/\bigl[\sqrt{\tilde h(z_h)}\,\tilde\alpha(z_h)\bigr]$, so fixing $T = |\tilde f'(z_h)|/(4\pi)$ fixes $\tilde\alpha(z_h)$ and hence $\tilde h'(z_h)$.
Deformations with $\delta h(z_h) = \delta h'(z_h) = 0$ --- such as $\delta h_2$ in Step~4 below --- therefore preserve the full set of horizon thermodynamic data, $s$ and $T$, while still changing the geometry.

Moreover, the boundary derivative $a \equiv \tilde f'(0) = \tilde h'(0)$ is a free parameter shared by $\tilde f$ and $\tilde h$.
To see this, differentiate~\eqref{eq:f_from_g} using $\tilde\alpha(0) = 1$, $\tilde r'(0) = 1$, and $\tilde\alpha'(0) = -\tilde h'(0)/2$:
\begin{equation}\label{eq:fprime0}
  \tilde f'(0) = g'(0) + \tilde h'(0)\,.
\end{equation}
For the Gubser--Rocha metric, $f_{\text{GR}}'(0) = h_{\text{GR}}'(0) = \tfrac{3}{2}Q$, giving $g'(0) = 0$.
Since $g'(0)$ is a property of $g(r)$ (fixed by the data), every member of the degenerate family satisfies $\tilde f'(0) = \tilde h'(0)$ with the common value $a$ unconstrained by $S_{EE}$.
This is the parameter that drifts during extended neural network training without Wilson loop data (figure~\ref{fig:a_drift}).

Thus the full set of constraints on $\tilde h$ is: $\tilde h(0) = 1$, $\tilde h(z_h) = z_h^2/r_h^2$, and --- if the temperature is also included in the data --- $\tilde h'(z_h)$: finitely many point conditions on a smooth function, with the boundary slope $a = \tilde h'(0)$ free. This leaves infinitely many degrees of freedom.

\paragraph{Step 4: Explicit examples.}
To illustrate the degeneracy concretely, consider deformations of the Gubser--Rocha warp factor $h_{\text{GR}}(z) = (1+Qz)^{3/2}$.
Define the one-parameter family
\begin{equation}\label{eq:h_deformed}
  \tilde h_\epsilon(z) = h_{\text{GR}}(z) + \epsilon\,\delta h(z)\,,
\end{equation}
where $\delta h(z)$ is any smooth function satisfying $\delta h(0) = 0$ and $\delta h(z_h) = 0$.
For example:
\begin{align}
  \delta h_1(z) &= z\,(z_h - z)\,, \notag\\
  \delta h_2(z) &= z^2\,(z_h - z)^2\,, \notag\\
  \delta h_3(z) &= \sin(\pi z/z_h)\,.
\end{align}
Each choice gives a different $\tilde h_\epsilon(z)$ satisfying $\tilde h(0) = h_{\text{GR}}(0) = 1$ and $\tilde h(z_h) = h_{\text{GR}}(z_h)$.
The function $\alpha_{\text{GR}}(z) = 1 - 3Qz/[4(1+Qz)]$ decreases monotonically from $1$ at the boundary to $\alpha_{\text{GR}}(z_h) = 1 - 3Q/[4(1+Q)]$, equal to $5/8$ for $Q = 1$, so $\tilde\alpha > 0$ holds for sufficiently small $|\epsilon|$ --- and in fact for a large range.
For $\delta h_1 = z(1-z)$ and $Q = 1$ the positivity condition $2\tilde h - z\tilde h' > 0$ reduces to $\sqrt{1+z}\,(2 + z/2) + \epsilon\,z > 0$ on $[0,1]$, so $\tilde\alpha > 0$ if and only if $\epsilon > -5\sqrt{2}/2 \approx -3.54$; in particular \emph{every} $\epsilon \geq 0$ yields an admissible deformation.
The corresponding $\tilde f_\epsilon$ from~\eqref{eq:f_from_g} produces identical $S_{EE}(l)$ for each such~$\epsilon$, yet the spacetime geometry is different in each case.
This demonstrates that neither the boundary values of~$h$ nor the thermal entropy suffice to resolve the degeneracy; the choice $\delta h_2$, for which $\delta h(z_h) = \delta h'(z_h) = 0$, preserves the temperature as well.

\paragraph{Step 5: UV expansion.}
Higher-order UV data does not resolve the degeneracy either; in fact, the flat direction admits an explicit order-by-order parametrization.
Expanding $f = 1 + az + \sum_{n\geq 2} f_n z^n$ and $h = 1 + az + \sum_{n\geq 2} h_n z^n$ (where $a = f'(0) = h'(0)$ is the shared boundary derivative), a direct series computation of $g = \alpha^2 f$ in the Bilson coordinate gives
\begin{equation}\label{eq:g2_expansion}
  g(r) = 1 + \left(f_2 + \tfrac{1}{4}a^2 - 2h_2\right) r^2 + \left(f_3 - 3h_3\right) r^3 + O(r^4)\,,
\end{equation}
with no linear term, consistent with $g'(0) = 0$ for a shared boundary slope.
The structure visible in~\eqref{eq:g2_expansion} is general: at every order,
\begin{equation}\label{eq:gn_structure}
  g_n = f_n - n\,h_n + P_n\!\left(a,\, f_2, \ldots, f_{n-1},\, h_2, \ldots, h_{n-1}\right),
\end{equation}
with $P_n$ a polynomial in the lower-order coefficients.
The linear structure follows by varying $g = \alpha^2 f$ at fixed lower orders: a variation $\delta f_n\, z^n$ enters $g_n$ directly with unit coefficient; a variation $\delta h_n\, z^n$ enters through $\alpha^2$ with coefficient $-n$, while its effect through the coordinate change $r = z/\sqrt{h}$ is $g'(r)\,\delta r = O(r^{n+2})$ and does not contribute at order~$n$.
We have verified~\eqref{eq:gn_structure}, including the explicit coefficients in~\eqref{eq:g2_expansion}, by symbolic computation through $O(r^6)$.
Consequently, each order of the UV expansion supplies one datum, $g_n$, for the two new coefficients $(f_n, h_n)$: only the combination $f_n - n h_n$ is determined, while the orthogonal direction
\begin{equation}\label{eq:kernel}
  \delta f_n = n\,\delta h_n
\end{equation}
is invisible to the entanglement data at every order --- the $n = 1$ case being precisely the shared boundary derivative~$a$, whose unconstrained drift is exhibited in figure~\ref{fig:a_drift}.

As shown in section~\ref{sec:bilson_hashimoto}, supplementing the entanglement data with Wilson loop data $V(L)$ breaks this degeneracy completely, determining both $f$ and $h$ to high precision.

\section{Boundary asymptotics: CFT and bulk perspectives}
\label{app:asymptotics}

The network encodings of sections~\ref{sec:bc} and~\ref{sec:inverse_gr} impose a definite near-boundary structure: the endpoint behavior $z \sim (l/2 - x)^{1/d}$ of the surface ansatz, the boundary values $f(0) = h(0) = 1$, the horizon condition $f(z_h) = 0$, and the UV parametrization~\eqref{eq:Vmodel_encoding}.
One may worry that this built-in structure secretly feeds part of the answer into the reconstruction.
The claim of this appendix is that it does not, because none of it is specific to the geometry being sought: every encoded item is common to \emph{all} candidate geometries, and corresponds to information about the boundary theory that is available before any reconstruction is attempted.
Each item can be read in two equivalent ways, and we present both.
From the boundary (CFT) side, the small-width behavior of the observables is fixed by conformal symmetry alone; the boundary values of the metric express the existence of an ultraviolet fixed point; the horizon condition expresses that the state is thermal; and the powers of $z$ in the metric parametrization reflect the dimensions of the operators of the dual theory.
From the bulk side, the same statements follow because every geometry in the class approaches pure AdS near the boundary, so the near-boundary shape of trial surfaces and trial metrics is identical for every candidate; this universality is checked numerically below on the nontrivial reconstructed backgrounds.
What the encodings never fix is the interior of $f$ and $h$ --- the part that the boundary data must determine.

\paragraph{CFT perspective: the observables.}
For a strip of width $l$ the only scale in the entanglement entropy is $l$ itself, so conformal invariance of the UV fixed point fixes the small-$l$ behavior up to a constant: beyond the area-law divergence, the finite part scales as $\Delta S_{EE} \propto V_\perp/l^{d-2}$~\cite{Ryu:2006bv,Ryu:2006ef}.
In $d = 2$ the corresponding statement is the twist-operator result, with known twist dimension $\Delta_n = \frac{c}{12}\left(n - \frac{1}{n}\right)$~\cite{Calabrese:2004}; in $d > 2$ the twist fields become codimension-two defect operators, and only scale invariance is needed to fix the power above.
Likewise, conformal invariance of the Wilson-line defect fixes the small-$L$ quark--antiquark potential to the Coulomb form $V(L) \propto -1/L$~\cite{Maldacena:1998im}.
These statements constrain the ultraviolet behavior of the boundary \emph{data}; they involve no dynamical input about the bulk interior.

\paragraph{Bulk perspective: the anchoring exponents.}
Every metric in the classes~\eqref{eq:metric} and~\eqref{eq:metric_general} approaches pure AdS as $z \to 0$: $f, h \to 1$.
Because the strip preserves translations in $x$, the area and Nambu--Goto functionals possess conserved Hamiltonians --- \eqref{eq:firstintegral}, \eqref{eq:hamiltonian_gr} and~\eqref{eq:wl_hamiltonian} --- a Noether statement that requires no boundary-value problem to be solved.
Evaluating them near the boundary, where $f, h \to 1$, the interior functions drop out.
For the RT surface in the class~\eqref{eq:metric} one finds $|z'| \simeq (z_*/z)^{d-1}$, hence
\begin{equation}\label{eq:endpoint_rt}
  z(x) \simeq \left[d\, z_*^{d-1}\left(\tfrac{l}{2} - x\right)\right]^{1/d},
\end{equation}
and the warp factor $h$ in the class~\eqref{eq:metric_general} modifies only the prefactor.
For the Nambu--Goto string, \eqref{eq:wl_hamiltonian} gives $|z'| \simeq \hat z_*^2/(z^2\sqrt{f_* h_*})$, hence
\begin{equation}\label{eq:endpoint_ng}
  z(x) \simeq \left[\frac{3\,\hat z_*^{2}}{\sqrt{f_* h_*}}\left(\tfrac{L}{2} - x\right)\right]^{1/3},
\end{equation}
with exponent $1/3$ in \emph{any} boundary dimension; it coincides with the RT exponent $1/d$ precisely for $d = 3$, which is why the same encoding serves both the L-model and the W-model in section~\ref{sec:inverse_gr}.
Because every metric in the class approaches pure AdS near the boundary, the same reasoning connects these anchoring exponents to the CFT scalings above.
As $L \to 0$ the string turning point scales as $\hat z_* \propto L$, so the entire worldsheet is confined to the near-boundary region, where every metric of the class reduces to pure AdS.
Scale invariance then forces the profile to be self-similar, $z(x) = L\,w(x/L)$ with a universal shape function $w$, and evaluating the Nambu--Goto action on such a profile gives $V_{\rm reg}(L) \propto -1/L$ to leading order --- the bulk realization of the Coulomb law fixed by conformal invariance on the boundary.
The endpoint behavior~\eqref{eq:endpoint_ng} is the local expansion of the same self-similar profile near its anchor, and it persists at \emph{all} $L$ because the anchor region $z \to 0$ remains in the pure-AdS regime even when the turning point probes the interior.
The identical argument applied to the RT surface connects the exponent~\eqref{eq:endpoint_rt} with the scaling $\Delta S_{EE} \propto V_\perp / l^{d-2}$.
As a numerical confirmation, fitting $\gamma = d\log z / d\log(l/2 - x)$ on exact profiles in the window $z \in [10^{-5}, 10^{-3}]$ gives:
\begin{center}
\begin{tabular}{lccc}
\toprule
Profile & Pure AdS & Nontrivial background & Universal value \\
\midrule
RT surface, $d=4$ & $0.25000$ & $0.25000$ (AdS-Schwarzschild) & $1/d = 0.25$ \\
RT surface, $d=3$ & $0.33333$ & $0.33338$ (Gubser--Rocha, $Q{=}1$) & $1/d = 1/3$ \\
NG string, $d=3$ & $0.33333$ & $0.33338$ (Gubser--Rocha, $Q{=}1$) & $1/3$ \\
\bottomrule
\end{tabular}
\end{center}
We use the conserved Hamiltonians here only to exhibit the universality of the exponents; the reconstruction itself never invokes them.

\paragraph{CFT perspective: the metric parametrization.}
The conditions $f(0) = h(0) = 1$ define the asymptotically AdS class --- the statement that the dual theory has an ultraviolet fixed point --- while $f(z_h) = 0$ encodes thermality of the state at a temperature that is boundary data.
The shared boundary slope $f'(0) = h'(0)$ is not assumed but \emph{derived from the entanglement data} in appendix~\ref{app:degeneracy}.
The remaining structure of the parametrization~\eqref{eq:Vmodel_encoding} --- which powers of $z$ appear --- reflects the operator content of the dual theory: a bulk scalar dual to an operator of dimension $\Delta$ falls off as $z^{d-\Delta}$ and $z^{\Delta}$, feeding the metric functions at the corresponding orders.
For the AdS$_4$ Gubser--Rocha model used here, the dilaton of the underlying Einstein--Maxwell--dilaton action~\cite{Gubser:2009qt,Ahn:2023} has mass $m^2 = -2$, obtained by expanding its potential about the AdS vacuum, so it behaves as $\varphi \propto z$ near the boundary and is dual to an operator of dimension $\Delta = 2$ (or $\Delta = 1$ in the alternative quantization), from $\Delta(\Delta - 3) = m^2$; the integer-power structure of~\eqref{eq:Vmodel_encoding} presumes precisely this integer operator dimension.
In this sense the ansatz encodes spectroscopic data of the boundary theory rather than properties of the particular bulk solution.

We note that all of the asymptotic statements above can equivalently be obtained by solving the field equations in the near-boundary region, where the geometry reduces to pure AdS.
The substantive point is that none of them involves the dynamics of the bulk \emph{interior}, which is precisely what the reconstruction determines from the data.

\section{Wilson loop derivation}
\label{app:wilson}

We derive the holographic Wilson loop potential and the derivative relation used in section~\ref{sec:bilson_hashimoto}.
A rectangular Wilson loop of temporal extent~$\mathcal{T}$ and spatial separation~$L$ in the metric~\eqref{eq:metric_general} is computed by a string worldsheet with $t = \tau$, $x = \sigma$, $z = z(x)$.
The induced metric has components $G_{\tau\tau} = -f/z^2$ and $G_{\sigma\sigma} = h/z^2 + z'^2/(z^2 f)$, with determinant $-\det G = (fh + z'^2)/z^4$.
The Nambu--Goto action gives the potential
\begin{equation}\label{eq:wl_induced}
  V(L) = \frac{1}{2\pi\alpha'}\int_{-L/2}^{L/2}dx\;\frac{1}{z^2}\sqrt{f\,h + z'^2}\,.
\end{equation}
Note that the combination $fh$ appears under the square root, in contrast to $h^2 + hz'^2/f$ for the RT area.

Since the Lagrangian has no explicit $x$-dependence, the Hamiltonian is conserved:
\begin{equation}\label{eq:wl_hamiltonian}
  \mathcal{H}_W = -\frac{f\,h}{z^2\sqrt{fh + z'^2}} = -\frac{\sqrt{f_*\,h_*}}{\hat z_*^2}\,,
\end{equation}
where $\hat z_*$ denotes the turning point of the string profile (distinct from the RT turning point $z_*$).
Solving for $z'$ and integrating yields the half-separation
\begin{equation}\label{eq:wl_L}
  \frac{L}{2} = \int_0^{\hat z_*}\frac{dz}{\sqrt{fh}\;\sqrt{\frac{fh\,\hat z_*^4}{z^4\,f_*h_*} - 1}}\,.
\end{equation}
The regularized potential (after subtracting the energy of two straight strings from boundary to horizon) is
\begin{equation}\label{eq:wl_Vreg}
  V_{\text{reg}}(L) = \frac{1}{\pi\alpha'}\!\left[\int_0^{\hat z_*}\!\frac{1}{z^2}\!\left(\frac{1}{\sqrt{1 - \frac{z^4 f_*h_*}{\hat z_*^4 fh}}} - 1\right)\!dz - \int_{\hat z_*}^{z_h}\!\frac{dz}{z^2}\right].
\end{equation}

The derivative of the regularized potential with respect to the separation is obtained by differentiating $L(\hat z_*)$ and $V_{\text{reg}}(\hat z_*)$ via the Leibniz rule.
The endpoint divergences cancel in the ratio, yielding
\begin{equation}\label{eq:dVdL_app}
  \frac{dV_{\text{reg}}}{dL} = \frac{1}{2\pi\alpha'}\,\frac{\sqrt{f_*\,h_*}}{\hat z_*^2}\,.
\end{equation}
In the Bilson--Hashimoto reconstruction of section~\ref{sec:bilson_hashimoto}, this identity provides the parametric variable $h_0 = 2\pi\alpha'\,dV_{\text{reg}}/dL = \sqrt{f_*h_*}/\hat z_*^2 = \sqrt{F_0 G_0}$ from boundary data.

\section{Semi-analytical reconstruction of the Gubser--Rocha metric}
\label{app:numerics}

This appendix presents the complete semi-analytical reconstruction of the Gubser--Rocha metric from the boundary data $S_{EE}(l)$ and $V(L)$ alone, combining the Bilson inversion~\cite{Bilson:2008,Bilson:2010ff} for the spatial metric with the Hashimoto inversion~\cite{Hashimoto:2020} for the timelike component.
The reconstruction proceeds entirely in the Bilson radial coordinate~$r$ and determines the two independent metric functions $g(r)$ and $\chi(r)$ defined in~\eqref{eq:metric_r}.
All integrals are evaluated with adaptive Gauss--Kronrod quadrature in double precision.

\paragraph{Step 1: Bilson inversion $S_{EE}(l) \to g(r)$.}

The Bilson formula~\eqref{eq:bilson_inversion}~\cite{Bilson:2008,Bilson:2010ff} takes $l(r_*)$ as input and returns $g(r)$.
The Bilson coordinate $r_*$ of the RT turning point is extracted directly from the entanglement entropy data via the derivative identity $dS_{EE}/dl = \Omega/(4G_N\,r_*^2)$~\cite{Ahn:2024}, which gives $r_*(l) = 1/\sqrt{d\tilde S/dl}$ where $\tilde S \equiv (4G_N/\Omega)\,S_{EE}$.
Following ref.~\cite{Ahn:2024}, a smooth power-law fit to $\tilde S(l)$ is used before differentiating.
At small $r_*$ the data is supplemented with the pure-AdS asymptotics $l(r_*) \approx 2r_*\sqrt{\pi}\,\Gamma(3/4)/\Gamma(1/4)$.

The Bilson integral~\eqref{eq:bilson_inversion} is evaluated via the substitution $u = (r_*/r)^4$, and both this integral and the turning-point integrals used to generate the input data are regularized with the further substitution $u = \sin^2\theta$, which removes all endpoint singularities and allows the quadrature to reach machine precision.
Spline differentiation of the Bilson integral yields $g(r)$ with a median relative error of $6 \times 10^{-9}$.

\paragraph{Step 2: Coordinate change to Hashimoto form.}

Starting from the Bilson metric~\eqref{eq:metric_r} with the now-known $g(r)$, we define a new radial coordinate
\begin{equation}\label{eq:eta_def_app}
  \eta(r) = \ln r + \int_0^r\left[\frac{1}{r'\sqrt{g(r')}} - \frac{1}{r'}\right]dr'\,.
\end{equation}
The subtraction of the pure-AdS integrand $1/r'$ ensures convergence at $r = 0$ and eliminates numerical cancellation near the boundary.
In $\eta$-coordinates the metric takes the Hashimoto form~\cite{Hashimoto:2020}:
\begin{equation}\label{eq:metric_eta_app}
  ds^2 = -F(\eta)\,dt^2 + G(\eta)\!\left(dx^2 + dy^2\right) + d\eta^2\,,
\end{equation}
with $G(\eta) = 1/r(\eta)^2$ (known) and $F(\eta) = g\,e^{-\chi}/r^2$ (unknown, contains $\chi$).

\paragraph{Step 3: Wilson loop inversion $V(L) \to \chi(r)$.}

Since~\eqref{eq:metric_eta_app} is precisely the metric form treated by Hashimoto~\cite{Hashimoto:2020}, we apply his non-zero-temperature inversion directly.
The Wilson loop derivative identity $h_0 = 2\pi\alpha'\,dV_{\rm reg}/dL = \sqrt{F_0\,G_0}$ (cf.~section~\ref{sec:inverse_gr}) provides the parametric variable $h_0$ from boundary data; inverting gives $L(h_0)$.
Hashimoto's Abel inversion then determines the structure function
\begin{equation}\label{eq:sigma_app}
  \sigma(H) = \frac{-1}{\pi}\;\frac{d}{dH}\int_H^{\infty}\frac{L(h_0)}{\sqrt{h_0^2 - H^2}}\,dh_0\,,
\end{equation}
where $H \equiv \sqrt{F\,G}$.
The equation relating $\eta$ and $H$,
\begin{equation}\label{eq:deta_dH_app}
  \frac{d\eta}{dH} = -\sigma(H)\,\sqrt{G(\eta)}\,,
\end{equation}
is separable, yielding
\begin{equation}\label{eq:separable}
  \Phi(\eta) \equiv \int_{-\infty}^{\eta}r(\eta')\,d\eta' = \int_H^{\infty}\sigma(H')\,dH' \equiv \Psi(H)\,.
\end{equation}
Both $\Phi$ and $\Psi$ are integrals of known functions.
Matching $\Phi(\eta) = \Psi(H)$ and inverting gives $H(\eta)$, hence $F = H^2/G$ and
\begin{equation}\label{eq:chi_from_FG}
  \chi(r) = \log\!\left[\frac{g(r)}{r^2\,F(\eta(r))}\right].
\end{equation}

\paragraph{Numerical implementation.}
\label{app:numerics_detail}

Several techniques are essential for achieving high precision:

\begin{enumerate}
\item \emph{$\theta$-substitution.}
All turning-point integrals for $l(z_*)$, $A_{\rm reg}(z_*)$, $L(z_*)$, $V_{\rm reg}(z_*)$ use the substitution $t = \sin^2\theta$ (after the initial $t = (z/z_*)^2$ mapping), which removes both endpoint singularities and produces smooth integrands on $[0,\,\pi/2]$.
The same substitution is applied to the Bilson integral and the Abel integral in~\eqref{eq:sigma_app} (via $h_0 = H/\cos\theta$).

\item \emph{AdS subtraction.}
Near the boundary, $g \to 1$, $\sigma \to 1/(2H^{3/2})$, and both $\Phi$ and $\Psi$ are dominated by the pure-AdS contributions $e^{\eta}$ and $1/\sqrt{H}$ respectively.
To avoid numerical cancellation, we compute the \emph{deviations}
\begin{equation}\label{eq:subtraction}
  \delta\Phi(\eta) = \Phi(\eta) - e^{\eta}\,,\qquad
  \delta\Psi(H) = \Psi(H) - \frac{1}{\sqrt{H}}\,,
\end{equation}
by integrating $r(\eta') - e^{\eta'}$ and $\sigma(H') - 1/(2H'^{3/2})$ respectively.
The pure-AdS parts cancel exactly in the matching $\Phi = \Psi$, and only the small deviations need to be resolved numerically.
This technique improves the precision of~$\chi$ from ${\sim}\,10^{-3}$ (without subtraction) to ${\sim}\,2 \times 10^{-5}$ (with subtraction).

\item \emph{UV tail.}
For $H$ beyond the data range, $L(h_0) \approx c/\sqrt{h_0}$ with $c$ determined from the largest available $h_0$ values.
The corresponding tail contributions to $\Psi$ are computed analytically.
\end{enumerate}

\paragraph{Results.}

The reconstruction achieves a median relative error of $6 \times 10^{-9}$ on $g(r)$ (from $S_{EE}$ alone) and a median absolute error of $3 \times 10^{-6}$ on $\chi(r)$ (from the combination of $S_{EE}$ and $V(L)$).
The ratio $f/h = g\,e^{-\chi}$, which characterizes the Lorentzian structure of the metric, is recovered to $3 \times 10^{-6}$ median relative error.
The results are shown in figure~\ref{fig:reconstruction} in the main text.

\section*{Acknowledgments}

This work was supported by the Bulgarian NSF grant KP-06-N88/1.
I acknowledge the open-source Get Physics Done (GPD) project~\cite{GPD} by Physical Superintelligence, whose AI-assisted physics research workflow (powered by Claude Opus 4.6 and Gemini 3.1) was helpful in carrying out aspects of this work.

\paragraph{Open Access.}
This article is distributed under the terms of the Creative Commons Attribution License (\href{https://creativecommons.org/licenses/by/4.0/}{CC-BY 4.0}), which permits any use, distribution and reproduction in any medium, provided the original author(s) and source are credited.

\end{document}